\documentclass[fleqn,usenatbib]{mnras}
\usepackage{newtxtext,newtxmath}
\usepackage[T1]{fontenc}
\usepackage{adjustbox}
\DeclareRobustCommand{\VAN}[3]{#2}
\let\VANthebibliography\thebibliography
\def\thebibliography{\DeclareRobustCommand{\VAN}[3]{##3}\VANthebibliography}
\graphicspath{{figures/}}

\usepackage{graphicx}	% Including figure files
\usepackage{amsmath,bm}	% Advanced maths commands
\usepackage{wasysym}    % Extra astronomy symbols
\usepackage{xspace}
\usepackage[version=4]{mhchem}
\makeatletter % Workaround to only color the year for cite commands
  \patchcmd{\NAT@citex}
    {\@citea\NAT@hyper@{%
      \NAT@nmfmt{\NAT@nm}%
      \hyper@natlinkbreak{\NAT@aysep\NAT@spacechar}{\@citeb\@extra@b@citeb}%
      \NAT@date}}
    {\@citea\NAT@nmfmt{\NAT@nm}%
    \NAT@aysep\NAT@spacechar\NAT@hyper@{\NAT@date}}{}{}

  \patchcmd{\NAT@citex}
    {\@citea\NAT@hyper@{%
      \NAT@nmfmt{\NAT@nm}%
      \hyper@natlinkbreak{\NAT@spacechar\NAT@@open\if*#1*\else#1\NAT@spacechar\fi}%
        {\@citeb\@extra@b@citeb}%
      \NAT@date}}
    {\@citea\NAT@nmfmt{\NAT@nm}%
    \NAT@spacechar\NAT@@open\if*#1*\else#1\NAT@spacechar\fi\NAT@hyper@{\NAT@date}}
    {}{}
\makeatother

\newcommand\Msun{\text{M}_{\astrosun}} % requires the wasysym package
\newcommand\HI{\ion{H}{I}\xspace} % neutral hydrogen
\newcommand\HII{\ion{H}{II}\xspace} % ionized hydrogen
\newcommand\HeI{\ion{He}{I}\xspace} % neutral helium
\newcommand\HeII{\ion{He}{II}\xspace} % singly ionized helium
\newcommand\HeIII{\ion{He}{III}\xspace} % doubly ionized helium
\newcommand\arepo{\textsc{arepo}\xspace}
\newcommand\areport{\mbox{\textsc{arepo-rt}}\xspace}

\newcommand\orcid[1]{\href{http://orcid.org/#1}{\adjustbox{trim={-.15\width} {0\height} {-.15\width} {0\height},clip}{\includegraphics[height=12pt]{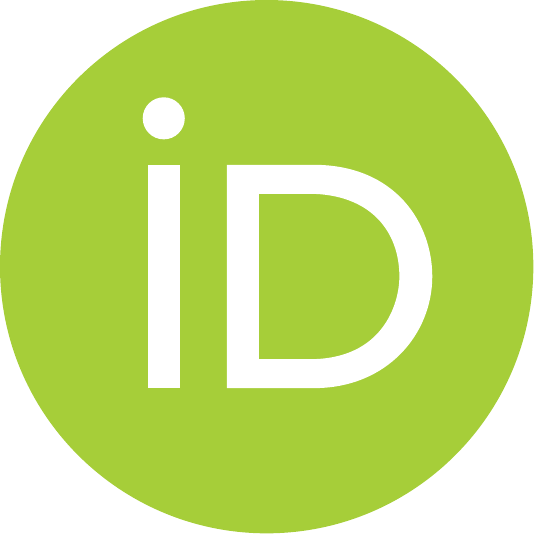}}}}

\title[Simulating ionization feedback]{Simulating ionization feedback from young massive stars: \\ impact of numerical resolution}

\author[Y. Deng et al.]{%
Yunwei Deng\orcid{0000-0002-7478-6427}$^{1,2,3}$,
Hui Li\orcid{0000-0002-1253-2763}$^{1,4}$\thanks{E-mail: \href{mailto:hliastro@tsinghua.edu.cn}{hliastro@tsinghua.edu.cn}},
Rahul Kannan\orcid{0000-0001-6092-2187}$^{5}$,
Aaron Smith\orcid{0000-0002-2838-9033}$^{6}$,
Mark Vogelsberger\orcid{0000-0001-8593-7692}$^{2,7}$
\newauthor
and Greg L. Bryan\orcid{0000-0003-2630-9228}$^{4}$%}
\\%
$^{1}$Department of Astronomy, Tsinghua University, Beijing 100084, People’s Republic of China\\%
$^{2}$Department of Physics, Kavli Institute for Astrophysics and Space Research, Massachusetts Institute of Technology, Cambridge, MA 02139, USA\\%
$^{3}$School of Astronomy and Space Science, Nanjing University, Nanjing 210093, People’s Republic of China\\%
$^{4}$Department of Astronomy, Columbia University, New York, NY 10027, USA\\%
$^{5}$Department of Physics and Astronomy, York University, 4700 Keele Street, Toronto, ON M3J 1P3, Canada\\%
$^{6}$Department of Physics, The University of Texas at Dallas, Richardson, Texas 75080, USA\\%
$^{7}$The NSF AI Institute for Artificial Intelligence and Fundamental Interactions, Massachusetts Institute of Technology, Cambridge, MA 02139, USA%
}
\date{Accepted XXX. Received YYY; in original form ZZZ}

\pubyear{2023}
\begin{document}
\label{firstpage}
\pagerange{\pageref{firstpage}--\pageref{lastpage}}
\maketitle

\begin{abstract}
Modelling galaxy formation in hydrodynamic simulations has increasingly adopted various radiative transfer methods to account for photoionization feedback from young massive stars. However, the evolution of \HII regions around stars begins in dense star-forming clouds and spans large dynamical ranges in both space and time, posing severe challenges for numerical simulations in terms of both spatial and temporal resolution that depends strongly on gas density ($\propto n^{-1}$). 
In this work, we perform a series of idealized \HII region simulations using the moving-mesh radiation-hydrodynamic code \areport to study the effects of numerical resolution. 
The simulated results match the analytical solutions and the ionization feedback converges only if the Str\"omgren sphere is resolved by at least $10$--$100$ resolution elements and the size of each time integration step is smaller than $0.1$ times the recombination timescale.
Insufficient spatial resolution leads to reduced ionization fraction but enhanced ionized gas mass and momentum feedback from the \HII regions, as well as degrading the multi-phase interstellar medium into a diffuse, partially ionized, warm ($\sim8000$\,K) gas. On the other hand, insufficient temporal resolution strongly suppresses the effects of ionizing feedback. This is because longer timesteps are not able to resolve the rapid variation of the thermochemistry properties of the gas cells around massive stars, especially when the photon injection and thermochemistry are performed with different cadences. Finally, we provide novel numerical implementations to overcome the above issues when strict resolution requirements are not achievable in practice.
\end{abstract}


%%%%%%%%%%%%%%%%%%%%%%%%%%%%%%%%%%%%%%%%%%%%%%%%%%

%%%%%%%%%%%%%%%%% BODY OF PAPER %%%%%%%%%%%%%%%%%%
\section{Introduction}

Feedback from young massive stars plays a key role in the evolution of giant molecular clouds (GMCs), galaxies, and the intergalactic medium throughout the Universe. Born in their natal clouds, massive stars provide feedback to the ambient interstellar medium (ISM) with ultra-violet (UV) radiation and stellar winds, and die as core-collapse supernovae (SNe). Ionizing radiation, together with stellar winds, from these massive stars alters the ionization state of the ambient gas, and injects comparable thermal energy and momentum as SNe \citep[e.g.][]{2013ApJ...770...25A,2015MNRAS.448.3248G,2021MNRAS.505.3470J}. Such an energetic early feedback mechanism can disperse GMCs on a short time scale of $\sim 1.5$\,Myr \citep{2019Natur.569..519K} prior to the SN explosion \citep[e.g.][]{2010ApJ...710L.142F,2012MNRAS.424..377D,2014MNRAS.442..694D,2016ApJ...829..130R,2019MNRAS.487..364L} and thus is crucial to regulate the star formation in galaxies \citep[e.g.][]{2012MNRAS.421.3488H,2018ApJ...865L..22E,2020MNRAS.491.3702H,2022MNRAS.509..272C}. It is also a key ingredient of galaxy formation simulations and significantly affects predicted galactic properties \citep[e.g.][]{1986PASP...98.1014S,2004ApJ...610...14W,2014MNRAS.437.2882K,Kannan2020a,2022MNRAS.511.4005K,Kannan2022b,Garaldi2022,Smith2022}.

Recently, with the surging interest in the role of radiative feedback in the formation and evolution of galaxies, several radiation-hydrodynamics implementations have been developed \citep[e.g.][]{2008MNRAS.387..295A,2009MNRAS.396.1383P,2011MNRAS.414.3458W,2013MNRAS.436.2188R,2018MNRAS.475.2822J,2019MNRAS.485..117K,2020ApJ...905...27S,Chan2021,Peter2023}. Sophisticated numerical algorithms have significantly reduced the cost of solving the radiative transfer (RT) equations, making it more affordable for cosmological simulations. 
However, radiative feedback from individual massive stars originates from very small $\sim10^4$\,K fully ionized regions, namely \textit{initial} Str\"omgren spheres \citep[][]{1939ApJ....89..526S} . Typically, the mass of the initial Str\"omgren sphere (initial Str\"omgren mass) is only on the order of several solar masses and its formation occurs within a short timespan of hundreds of years. Driven by these ionized cores, compact \HII regions expand outward and deposit momentum and kinetic energy to the turbulent ISM until the massive stars die in a few million years. Capturing such small-scale physics requires high spatial and temporal resolution and thus tremendous computational resources, which is often impractical to treat directly in galaxy formation simulations. Therefore, modelling the impact of individual \HII regions (photo-heating and expansion) from massive stars is numerically challenging.

The tiny mass and rapid formation of the initial Str\"omgren spheres make them extremely challenging to model in galaxy formation simulations. The largest-volume cosmological simulations of galaxy formation can only afford mass resolutions larger than $10^6\,\Msun$ for baryonic cells (e.g. \citealt{2015MNRAS.446..521S}; \citealt{2018MNRAS.473.4077P}; \citealt{ 2022MNRAS.511.4005K, Kannan2022c};
\citealt{2022arXiv221010059H}, and see \citealt{2020NatRP...2...42V} for a recent review). Even zoom-in simulations of single Milky Way-like galaxies \citep[e.g.][]{2015MNRAS.451...34R,2018MNRAS.480..800H,2019MNRAS.489.4233M} usually have mass resolutions spanning from $10^3$--$10^5\,\Msun$ per gas resolution element, which is far larger than the initial Str\"omgren mass. Although recent dwarf galaxy simulations \citep[e.g.][]{2019MNRAS.482.1304E,2020MNRAS.491.1656A,2020ApJ...891....2L,2021MNRAS.501.5597G} have begun to treat feedback from individual stars in dwarf galaxies with a much higher mass resolution of several solar masses, directly resolving the individual \HII regions still remains a numerical issue. Scaling inversely with density, the Str\"omgren spheres are normally unresolved in gas with $n_\text{H} \gtrsim 10\,\text{cm}^{-3}$ for high-resolution galaxy radiation-hydrodynamic simulations \citep[e.g.][]{2015MNRAS.451...34R,Rosdahl2022}. Even for GMC simulations, resolving \HII regions can still be challenging in high-density regimes as both the mass and formation timescale of Str\"omgren spheres decrease significantly at high density ($\propto n^{-1}$).

Inadequate resolution often leads to problematic results in galaxy formation simulations. Historically, the inability to resolve the Sedov-Taylor phase of SN explosion leads to artificially rapid radiative cooling and forming too many stars too early in cosmological simulations
\citep{1992ApJ...391..502K}. This so-called {\it over-cooling} problem has been addressed by employing a variety of sub-grid prescriptions (see \citealt{{2017ARA&A..55...59N}} for a recent review). In terms of ionization feedback, several works have explored the consequences of insufficient resolution when simulating idealized \HII regions, including the impact on radial profiles and momentum feedback (e.g. \citealt{2011MNRAS.414.3458W}; \citealt{2015MNRAS.453.1324B}; \citealt{2021MNRAS.507..858P}; \citealt{2022MNRAS.510.2797P}; see also \citealt{Ivkovic2023}). However, the physics behind these numerical issues needs to be further clarified, especially in the context of specific RT methods and applications. Furthermore, most studies focus solely on spatial resolution and disregard the influence of temporal resolution. Therefore, there is an urgent need for well-understood solutions to address these resolution problems to enhance the convergence of current RT implementations in a more physical manner.

In this article, we perform a suite of radiation-hydrodynamic simulations of idealized \HII regions with the moment-based M1 closure RT implementation \citep{2019MNRAS.485..117K} in the moving-mesh code \arepo \citep{2010MNRAS.401..791S}. We study both the effects of spatial (mass) and temporal (time-stepping) resolution on the behaviour of simulated \HII regions. For simplicity, our tests are all performed in uniform pure hydrogen media, but the discussions and conclusions based on the hydrogen-only case are also valid for the other species by replacing the coefficients correspondingly. We address three issues arising from the insufficient spatial (mass) and temporal resolution on stellar ionization feedback: 
\begin{itemize}
    \item {\bf Insufficient spatial (mass) resolution}
\begin{enumerate}
    \item \textit{Over-ionization}: overestimation of the total ionized gas mass due to gas cells that can not be fully ionized, shifting the ionization-recombination balance to the ionization side;
    \item \textit{Over-heating}: overestimation of energy and momentum deposition due to the artificial heating of a substantial amount of partially ionized gas;
\end{enumerate}
    \item {\bf Insufficient temporal resolution}
\begin{enumerate}
    \item \textit{Missing photons}: underestimation of the overall feedback due to excess photons being absorbed without any effect on the ionization state of the gas.
\end{enumerate}
\end{itemize}
We also emphasize that {\it missing photons} can be a very serious problem when photon injection and thermochemistry have different cadences. We explain the physical reasons for these numerical issues and describe novel numerical implementations to solve these issues at an acceptable computational cost. 

This paper is organized as follows: We review the physics and analytic results of \HII region expansion in different phases in Section~\ref{sec:analytic}. In Section~\ref{sec:methods}, we describe our simulation setup and RT implementations. In Sections~\ref{sec:resolution} and \ref{sec:time-stepping}, we present the results of our tests for spatial and temporal resolution dependence, respectively. We then introduce %the possible
several solutions in Section~\ref{sec:corrections}. Lastly, we discuss and summarize our results in Section~\ref{sec:discussion}. 

\section{Anlytical theory of \HII regions}
\label{sec:analytic}
In this section, we review the physics of idealized \HII regions in a uniform medium. Because of the differences in the dominant physical processes and timescales, the evolution of an idealized \HII region around a massive star can be divided into two stages, namely the {\it formation} and {\it expansion} phase \citep[e.g.][]{1986ARA&A..24...49Y}. Realistic \HII regions expanding in the turbulent environment with a roughly power-law density profile \citep[e.g.][]{1977ApJ...214..488S,2003ApJ...585..850M,2018A&A...611A..88L} can expand preferentially toward the rarefied regions and launch a ``champagne flow" that rapidly ionizes and heats the surrounding cloud \citep{1990ApJ...349..126F,2019MNRAS.487.2200Z,2020MNRAS.492..915G,2021MNRAS.501.1352G}. Detailed discussions of \HII regions in realistic environments are outside the scope of this work on the impact of numerical resolution, we refer the reader to the aforementioned literature.

\subsection{Formation phase}
Once a hot massive star ignites in neutral gas, a so-called R(rarified)-type ionization front (I-front), characterized by a rapid expansion and a negligible density change across the I-front, is driven through the gas. This I-front leaves the gas behind the front hot ($\sim10^4$\,K) and ionized but otherwise almost undisturbed. The speed of this I-front is initially supersonic. It slows down gradually until the radius of the \HII region reaches the ``initial'' Str\"omgren radius. 

This idealized problem of a fully ionized, spherical region of uniform density with the ionization maintained by constantly emitted ionizing photons from a central massive star is known as a Str\"omgren sphere \citep{1939ApJ....89..526S}. Assuming the gas is pure hydrogen, a steady solution for the radius of the Str\"omgren sphere (Str\"omgren radius) can be found by requiring ionization equilibrium,
\begin{equation}
\label{equ:StromgrenRadius}
    R_\text{S} = \left( \frac{3Q}{4\pi n_\text{H}^2 \alpha_\text{B}} \right)^{1/3} \approx 0.315\,Q_{48}^{1/3}n_3^{-2/3}{\alpha_0}^{-1/3}\,\text{pc} \, ,
\end{equation}
where $\alpha_{\rm B}$ is the effective hydrogen radiative recombination rate for Case B approximation \citep[][]{1938ApJ....88...52B}, $\alpha_0 \equiv \alpha_{\rm B} / (2.59\times10^{-13}\,\text{cm}^3\text{s}^{-1})$, $Q$ is the rate of emission of ionizing photons with $Q_{48} \equiv Q / (10^{48}\,\text{ s}^{-1})$, and $n_\text{H}$
is the hydrogen density with $n_3 \equiv n_{\text{H}} / (10^3\,\text{cm}^{-3})$. 
Since the initial Str\"omgren sphere is almost fully ionized, the initial mass of ionized gas, \textit{initial Str\"omgren mass}, is
\begin{equation}
    M_\text{S} = \frac{4}{3}\pi R_\text{S}^3 n_\text{H} m_\text{H} = \frac{m_\text{H}}{\alpha_\text{B}} \frac{Q}{n_\text{H}} \approx 3.25\,Q_{48} n_3^{-1}{\alpha_0}^{-1}\,\Msun \, .
    \label{equ:StMass}
\end{equation}
Assuming the I-front is infinitely thin ($l_\text{mfp}\lesssim10^{-4}n_3^{-1}$\,pc), in the non-relativistic limit, the evolution of the initial Str\"omgren sphere in the formation phase follows \citep{1978ppim.book.....S}
\begin{equation}
    r_i = R_\text{S} \left(1 - e^{-t/t_\text{rec}} \right)^{1/3} \, ,
    \label{equ:FormEq}
\end{equation}
where $r_i$ is the radius of the I-front and $t_{\rm rec}$ is the recombination timescale given by
\begin{equation}
    \label{equ:trec}
    t_\text{rec} = \frac{1}{\alpha_\text{B} n_\text{H}} \approx 122.3\,n_3^{-1}{\alpha_0}^{-1} \,\text{yr}\, .
\end{equation}
Equations~(\ref{equ:StromgrenRadius}), (\ref{equ:StMass}), and (\ref{equ:trec}) show that the typical initial Str\"omgren spheres have sizes under a few parsecs, masses of several solar masses, and form within several hundreds of years. Such spatial and time scales are much smaller and shorter than the scales of the subsequent expansion phase in most astrophysical environments.

\subsection{Expansion phase}
Once the I-front reaches the initial  Str\"omgren radius ($R_\text{S}$), the high pressure of the ionized gas drives the expansion of the \HII region. It undergoes a rapid transition from R-type to D(dense)-type front: The ionized gas in photoionization equilibrium expands, driven by the pressure gradient relative to the neutral gas and causes a shock to separate from the I-front and proceed into the surrounding neutral gas. The neutral gas thus begins to accumulate between the shock front and the I-front. This expansion phase itself can also be divided into two phases: the \textit{early} phase, during which the pressure between the ionized gas and surrounding neutral gas is large; and a \textit{later} phase when the internal pressure asymptotically balances with the thermal pressure of the surrounding gas. 

Assuming a thin swept-up shell ($v_i\sim v_\text{shock}$) and applying the conservation of momentum flux across the I-front and shock front, \cite{1978ppim.book.....S} gave the classic solution of the time evolution of the expanding I-front. However, this solution deviates from the expansion at very early time because it ignores the inertia of shocked gas. Incorporated the inertia, \cite{2006ApJ...646..240H} obtained a solution which shows better agreement with the numerical
results \citep{2015MNRAS.453.1324B}:
\begin{equation}
\label{equ:H-I}
    r_{i} = R_\text{S} \left(1 + \frac{7}{4}\sqrt{\frac{4}{3}}\frac{c_i t}{R_\text{S}} \right)^{4/7} \, .
\end{equation}
As the Str\"omgren sphere expands into the uniform medium, its ionized mass continues to increase following a simple scaling relation \citep{2015MNRAS.453.1324B}:
\begin{equation}
\label{equ:Mi-t}
M_i(t) = M_\text{S} \left(\frac{r_i(t)}{R_\text{S}}\right)^{3/2} \, .
\end{equation}

\HII regions embedded in very dense ($\sim10^6$\,cm$^{-3}$) or warm ($>100$\,K) gas can enter their later phase of expansion \citep{2012MNRAS.419L..39R}. In the later phase, the expansion of the \HII region is stagnant while the pressure is equilibrated with the ambient gas so the classic analytic formulas are no longer valid. In hydrodynamic simulations, there is no clear boundary between the early and later phases because we follow the gas dynamics self-consistently. Since the momentum feedback from \HII regions is injected mostly in the early phase, we will therefore focus on the early phase expansion in this work.

\section{Methods}
In this section, we briefly describe the setup of the initial conditions and the terminology used to infer the definition of the mass and temporal resolution. We also recap the \areport implementation that is relevant for later discussions.
\label{sec:methods}

\subsection{Initial conditions}
To test the dependence of simulation results on the spatial and temporal resolution, we simulate idealized \HII regions formed by an individual massive star in uniform pure neutral hydrogen gas. This is a purely radiation-hydrodynamical test, i.e. gravity and magnetic fields are not included.  For each set of tests, a stellar particle with a steady ionizing photon emission rate $Q$ is placed at the centre of a box with size $L$ at least 3.6 times larger than the radius of the \HII region at the end of the simulations. The gas cells are initially arranged as a regular staggered mesh, which is constructed by overlapping two Cartesian meshes with a displacement of $0.45\Delta x$ along each axis from each other. Such a mesh configuration is adopted to stabilize the construction of the Voronoi tessellation and minimize density fluctuations in \arepo.

We initialize the gas cells as pure hydrogen to directly compare with analytical results introduced in Section~\ref{sec:analytic}. The initial temperature and density are different in our three tests, we will introduce each of them at the beginning of the corresponding section.

\subsubsection{Definitions of mass and temporal resolution}
\label{sec:resolutionDef}
Since the main goal of this paper is to explore the effects of numerical resolution on the ionization feedback of \HII regions, here we define explicitly the meaning of the mass and temporal resolution in our numerical experiments.

In quasi-Lagrangian codes like \arepo where the mass of gas cells are close to the target gas mass $M_\text{cell}$, we define the {\it mass resolution} $\mathcal{R}_i$ of the initial Str\"omgren sphere as the ratio between the initial Str\"omgren mass and the target gas mass:
\begin{equation}
    \mathcal{R}_i = \frac{M_\text{S}}{M_\text{cell}} = \frac{m_\text{H}}{\alpha_\text{B}}\frac{Q}{n_\text{H} M_\text{cell}} \, .
    \label{equ:resolution}
\end{equation}
Since the expansion of \HII regions is driven by the ionized gas inside the initial Str\"omgren sphere, $\mathcal{R}_i$ is critical to determine whether the Str\"omgren sphere and its feedback are resolved. 

The mass resolution has a direct connection to the physics of the evolution of \HII regions so it is much more useful and convenient than the spatial resolution in this work. Still, we define the conversion between mass and spatial resolution via
\begin{equation}
    \Delta x = \left(\frac{Q}{\alpha_{\rm B} n^2_\text{H}\mathcal{R}_i}\right)^{1/3} \, .
\end{equation}
For convenience, we use the terms `spatial resolution' and `mass resolution' interchangeably.

The {\it temporal resolution} is determined by the size of the time integration steps $\Delta t$. To examine how the results depend on temporal resolution, we fix the time steps to a given value in each run for all the cells, and both RT and hydro solvers in the main tests. In realistic simulations, gas cells and operations (e.g. hydrodynamics and RT) are allowed to have different time steps, enabling subcycling of processes with short characteristic timescales. For simplicity, we assume a uniform and fixed time step $\Delta t$ for all cells and operations throughout the simulation, except in Section~\ref{sec:Subcycling} and related sections. In these sections, we will illustrate how the temporal resolution issues can be notably amplified due to the discrepancy between the sizes of photon injection steps ($\Delta t_\star$) and RT/thermochemistry steps ($\Delta t_\text{RT}$).

\subsection{Radiative transfer implementation}
\label{sec:RT}
We use \areport \citep{2019MNRAS.485..117K}, the radiation-hydrodynamic extension of the moving-mesh hydrodynamic code \arepo \citep{2010MNRAS.401..791S} to handle the propagation of radiation and calculation of thermochemistry. \areport adopts a moment-based RT scheme with the M1 closure relation \citep{1984JQSRT..31..149L} to solve the set of hyperbolic conservation equations for photon density and flux. 
The UV radiation continuum is usually divided into three separate bins between
energy intervals of [13.6, 24.6, 54.4, $\infty$)\,eV mirroring the hydrogen and helium ionizing photon groups. With these three bins, $4\times3$ additional variables are stored in each cell to describe the volume density of photons $n_{\gamma}^i$ and the flux vector ${\bm F}_i$. The index $i$ here is the energy band corresponding to ionic species \HI, \HeI, and \HeII. For simplicity, these frequency-dependent coefficients are reduced to their effective value as only one single $>13.6$\,eV bin in our discussion if not specified. Moreover, the ionization state of each species in each cell is also stored as $x_\text{\HII}$, $x_\text{\HeII}$, and $x_\text{\HeIII}$, with $x_\text{\HI}$ and $x_\text{\HeI}$ set by detailed balancing. We adopt the on-the-spot approximation (OTSA) which assumes all recombination photons are re-absorbed immediately in the surrounding medium processes \citep[case~B,][]{2006agna.book.....O}. We note that although the OTSA may not capture the ionization profiles accurately in the optically thin regime near the source \citep{2014MNRAS.437.2816R}, it is a reasonable approximation in the high-density mostly neutral regions where the resolution issues occur (see Section~\ref{sec:where2correct}).

The RT equations are solved with an operator-splitting strategy over a pre-determined time-step $\Delta t$. Each RT step is decomposed into three sub-operations executed successively. Firstly, inject photons from the star into the neighbouring cells. Secondly, solve the RT equations and propagate photons in space. Lastly, solve the thermochemistry equations to determine the ionization states and conduct the heating and cooling processes in each cell. Since \arepo uses a two-stage second-order Runge–Kutta
integration scheme \citep[i.e. Heun's method, ][]{2016MNRAS.455.1134P}, the propagation and thermochemistry are in fact split into two halves within an RT step.

\subsubsection{Photon injection}
\label{sec:dumping}
In this work, we focus on the ionization feedback from an individual star. Each test or simulation contains a single {\it stellar particle} with a fixed ionizing photon emission rate $Q$ and ionization spectrum, which corresponds to an idealized massive star. For a steady stellar source, its ionizing photon emission rate $Q$ is divided into three frequency bins weighted by a given spectrum and $\sum_i Q_i=Q$. At time-step $\Delta t$, the number of injected photons from a stellar particle in the UV bin $i$ is $\Delta N_{\gamma}^i = Q_i\Delta t$. These $\Delta N_{\gamma}^i$ photons from the stellar particle are then weighted among a given number of its nearest neighbouring gas cells ($N_{\rm nb}=32$ by default). The weight factor $w_k$ for cell $k$ is calculated by the solid angle of the cell opened to the stellar particle. Therefore, the number of ionizing photons dumped into a cell $k$ close to a star at time-step $\Delta t$ is 
$\Delta N_{\gamma, k}^i = w_k \Delta N_{\gamma}^i$
where $\sum_k w_k=1$.

\subsubsection{Radiation transport}
\label{sec:transport}
\areport simulates the propagation of radiation by solving the set of hyperbolic conservation equations consisting of the zeroth- and first-order moments, which takes the form of 
\begin{align}
    &\frac{\partial n_{\gamma}^i}{\partial t} + \nabla\cdot{\bm F}_i = 0 \, , \notag \\
    &\frac{\partial {\bm F}_i}{\partial t} + \tilde{c}^2\nabla\cdot\mathbb{P}_i = 0 \, ,
\end{align}
where $\mathbb{P}_i$ is the pressure tensor related to $n_{\gamma}^i$ by the Eddington tensor.
The true speed of light is reduced to $\tilde{c}$ to prevent too small time-steps, which is known as the reduced speed of light approximation \citep{2001NewA....6..437G}. 
These equations are closed with the M1 closure relation.
There are two frequently-used schemes to evaluate the interface fluxes, the Harten-Lax-van Leer \citep[HLL,][]{harten1983upstream} flux function and the global Lax-Friedrich \citep[GFL,][]{rusanov1961calculation} function. The GFL function is more straightforward but makes the RT more diffusive, while the HLL function has its inherent directionality but makes isotropic radiation from stars asymmetric \citep{2013MNRAS.436.2188R}. For the sake of simplicity and symmetry, we adopt the GFL implementation in this work.

\subsubsection{Thermochemistry, cooling, and heating}
\label{sec:thermo}
In \areport, to calculate the number density of different ionic species in each cell, a series of equations are solved by a semi-implicit time integration approach based on the method outlined in \cite{2009MNRAS.396.1383P} after the photon injection and transport are finished in each step. Only if the internal energy changes by more than 10 per cent during a time-step, the equations will be solved implicitly by calling the functions from the SUNDIALS CVODE package \citep{hindmarsh2005sundials}. 

For the pure hydrogen case, as the simplest example, the  thermochemistry equation is
\begin{equation}
	\frac{\mathrm{d}n_\text{\HII}}{\mathrm{d}t}=-\alpha_\text{\HII}n_\text{\HII}n_{e}+\sigma_{e\text{\HI}}n_{e}n_\text{\HI}+\tilde{c}n_\text{\HI}\sum_i\sigma_{i\text{\HI}}n_{\gamma}^i
    \label{equ:thermoeq}
\end{equation}
where $n_\text{\HI}$, $n_\text{\HII}$, and $n_{\gamma}^i$ are the volume densities of neutral, ionized hydrogen, and photons in bin $i$, respectively, and $n_\text{\HI}=(1-x_\text{\HII})n_{\text{H}}$, $n_\text{\HII}=x_\text{\HII}n_{\text{H}}$, $n_{\gamma}^i = N_{\gamma}^i/V$. $\sigma_{i\text{\HI}}$ is the mean photoionization cross-sections for bin $i$, while $\sigma_{\text{e\HI}}$ and $\alpha_\text{\HII}$ are the collisional ionization cross-section and recombination rate taken from \cite{1996ApJS..105...19K}. The three terms on the RHS correspond to the number density rates for \HII recombinations, \HI collisional ionizations, and \HI photoionizations, respectively.

When conducting the heating and cooling, the internal energy per unit mass $u$ is stored in each cell. \areport calculates the heating ($\Gamma$) and cooling ($\Lambda$) rate per unit volume with parameters stored in each cell. The variation of thermal energy $\Delta u$ for a cell in a time-step $\Delta t$ is 
\begin{equation}
    \Delta u=\frac{1}{\rho}(\Gamma-\Lambda)\Delta t \, .
    \label{equ:DeltaU}
\end{equation}
The main heating process in the context of idealized \HII regions is photoionization heating by the deposited energy of the ionizing photons. The photoheating rates in a finite-frequency bins form can be given as
\begin{equation}
    \Gamma_j=n_j\sum_i\int_{\nu_{i1}}^{\nu_{i2}}\frac{4\pi J_{\nu}}{h\nu}\sigma_{j\nu}(h\nu-h\nu_{j})\,\mathrm{d}\nu \, ,
    \label{equ:phoheat}
\end{equation}
where $h$ is the Planck constant and $h\nu_{j}$ is the ionization potential of the ionic species $j$. The total photoheating rate can therefore be given as
$\Gamma=\sum_j\Gamma_j$.

We mainly consider the following cooling processes in the pure hydrogen \HII regions: recombination ($\Lambda_\text{rec}$), collisional excitation (bound-bound, $\Lambda_\text{bb}$), collisional ionization (bound-free, $\Lambda_\text{bf}$), and Bremsstrahlung (free-free, $\Lambda_\text{ff}$). The cooling equations we use follow \cite{1992ApJS...78..341C} and are listed in Appendix~\ref{sec:cooling}.
The total cooling rate is therefore calculated by
\begin{equation}
    \Lambda=\Lambda_\text{rec}+\Lambda_\text{bb}+\Lambda_\text{bf}+\Lambda_\text{ff} \, .
\end{equation}

\section{Effects of different mass (spatial) resolution}
\label{sec:resolution}
In this section, we study the resolution effects on the evolution of idealized \HII regions in their formation and expansion phases, respectively.

\subsection{Test 1 -- Formation phase and over-ionization}
\label{sec:Formation}

In the first test, we investigate the impact of resolution on the formation of the initial Str\"omgren sphere. To do this, we perform static Str\"omgren sphere tests by placing a steady source of hydrogen ionizing photons with a rate of $Q = 10^{63}$\,Gyr$^{-1}=3.17\times10^{46}$\,s$^{-1}$ (unity rate in \areport) in a uniform pure hydrogen neutral medium of number density $n_{\text{H}}=1$\,cm$^{-3}$. The effective \HI ionization cross-section $\sigma_\text{\HI}=3.0\times10^{-18}$\,cm$^{2}$. For this test, we turn off the cooling and heating of gas and the temperature is fixed at $T = 10^4$ K. The speed of light is reduced to $\tilde{c}=0.001c$. The chosen initial condition prioritizes simplicity, while the results can be scaled to any arbitrary initial conditions using the dimensionless resolution $\mathcal{R}_i$. To test the effect of mass resolution $\mathcal{R}_i$, we set a series of initial conditions with cell mass from about $10^{-3}\,\Msun$ to $10^{3}\,\Msun$, equivalent to $\mathcal{R}_i$ from $10^5$ to 0.1 or $\Delta x$ from $0.34$\,pc to $34$\,pc. 
All the cells are initially at rest and arranged in a regular staggered mesh. 

\begin{figure*}
    \includegraphics[width=2\columnwidth]{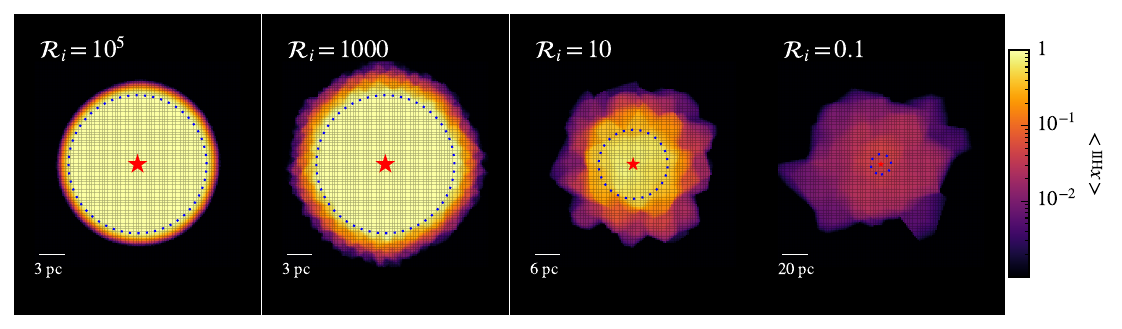}
    \caption{Structure of the Str\"omgren spheres (projected \HII fractions) in different mass resolution from the highest (${\cal R}_i=10^5$) to lowest (${\cal R}_i=0.1$) resolution simulations (Test~1, Section~\ref{sec:Formation}). The high-resolution Str\"omgren spheres are nearly spherical, demonstrating the ability of our code to obtain well-defined \HII regions with correct  Str\"omgren radius that is consistent with the analytical result as shown in the blue dotted circles. Notice that the spatial axes are rescaled for each map, while the \HII fractions are shown at the same scale.}
    \label{fig:nh2projections}
\end{figure*}

In Figure~\ref{fig:nh2projections}, we present the projected maps of the \HII fractions for simulations with different resolutions from ${\cal R}_i=10^5$ to ${\cal R}_i=0.1$ resolution with a spacing of $10^2$. The ${\cal R}_i=10^5$ and ${\cal R}_i=1000$ Str\"omgren spheres exhibit an ideal spherical morphology, indicating the ability of our code to capture the characteristic features of \HII regions. Compared with the analytic Str\"omgren radius ($R_\text{S}$, blue dotted circles), the high-resolution Str\"omgren spheres are almost fully ionized and concentrated inside $r\lesssim R_\text{S}$ with a sharp boundary. On the contrary, the low-resolution Str\"omgren spheres are much larger than $R_\text{S}$, filled with partly ionized gas with much lower \HII fractions.
 
\begin{figure}
    \includegraphics[width=\columnwidth]{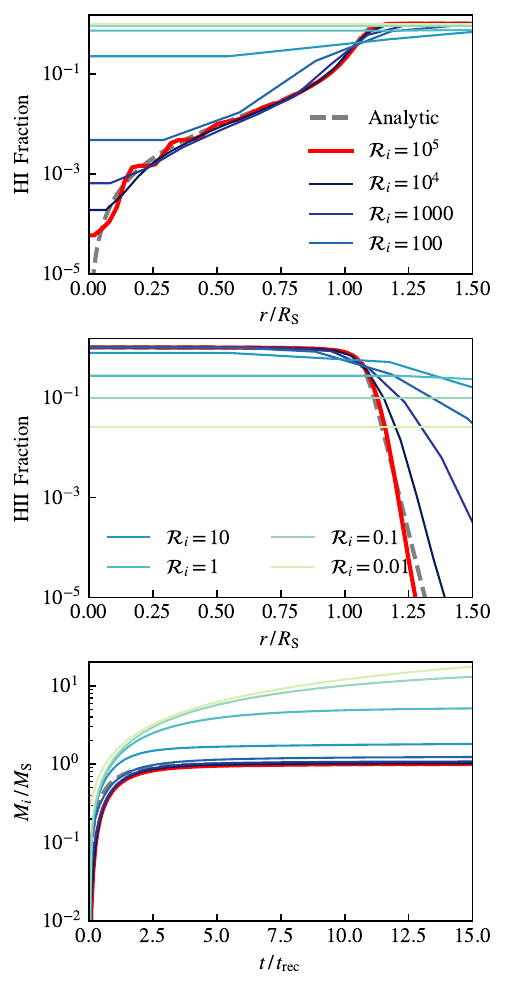}
    \caption{Results of the formation test (Test~1, Section~\ref{sec:Formation}). {\it Top (middle)} panels: profiles of neutral (ionized) fraction as a function of radius for different mass resolutions at $t \approx 15\,t_{\text{rec}}$; {\it Bottom} panel: evolution of ionized mass as a function of time for different mass resolutions.}
    \label{fig:FormationTest}
\end{figure}

In the top (middle) panels of Figure~\ref{fig:FormationTest} we show the steady neutral (ionized) fraction profiles at $t \approx 15\,t_{\text{rec}}$. The profile of the ionization fraction can be described by \citep{2006agna.book.....O}
\begin{equation}
\label{equ:OF06}
    \frac{x_\text{\HI}(r)}{4\pi r^2} Qe^{-\tau(r)}\sigma_\text{\HI} = x_\text{\HII}^2(r)n_{\text{H}}\alpha_\text{\HII} \, ,
\end{equation}
where $x_\text{\HI}(r) = n_\text{\HI}(r)/n_{\text{H}}$, $x_\text{\HII}(r)=1-x_\text{\HI}(r)$, the optical depth $\tau(r) = n_\text{H}\sigma_\text{\HI}\int_0^r x_{\HI}(r'){\rm d}r'$, and the cross-section $\sigma_{\HI}$ here is $3\times10^{-18}$\,cm$^{2}$. This analytic solution is plotted in Figure~\ref{fig:FormationTest} with the grey dashed curves as a benchmark. The analytic profile is well-reproduced only in the $\mathcal{R}_i\geq10^5$ simulation. With decreasing resolution, the profile becomes more and more smooth, the fractional ionized region becomes more extended while the fully ionized core region shrinks. In high-resolution cases, the ionization fraction $x_\text{\HII}$ reaches unity in the fully ionized core and there is no significant difference until about $\mathcal{R}_i=10$. However, once the \HII region transitions to being marginally resolved and then unresolved, the situation is very different. The effective size of one cell is larger than the initial Str\"omgren sphere so that the total ionized mass is only a small portion of the mass of one cell. Consequently, the ionization fraction is reduced to $\sim M_\text{S}/M_\text{cell}={\cal R}_i$. As we presented in Figure~\ref{fig:FormationTest}, $x_\text{\HII}$ falls down quickly to $\lesssim0.1$ in the $\mathcal{R}_i=0.1$ case.

Although the ionization fractions are averaged down and ionization profiles are smoothed away, the total mass of ionized gas $M_i$ is not conserved during the resolution smoothing. Assuming the Str\"omgren sphere is fully ionized, the time evolution of the I-front follows equation~(\ref{equ:FormEq}) and the total ionized mass can be described as
\begin{equation} \label{equ:MiEvo}
    M_{i}(t)=M_{\rm S}\left(1-e^{-t/t_{\text{rec}}}\right) \, .
\end{equation}
In the bottom panel of Figure~\ref{fig:FormationTest}, we present the evolution of $M_i$ for different resolutions $\mathcal{R}_i$, comparing with the analytic result of equation~(\ref{equ:MiEvo}). There is no significant difference among the $\mathcal{R}_i=10^5$, $10^4$, and $10^3$ simulations, and the initial Str\"omgren sphere can be regarded as well-revolved when the mass resolution is larger than $100$, at least from the perspective of the ionized mass. However, the total mass of ionized gas is enhanced in lower-resolution cases. In marginally resolved cases like $\mathcal{R}_i=10$ or $1$, the equilibrium mass of ionized gas is slightly larger than the actual mass, while in the unresolved case of $\mathcal{R}_i=0.1$ it is more than ten times larger. We refer to such overestimation of total ionized gas mass in low resolution cases as {\it over-ionization}. The time for the Str\"omgren sphere to reach a steady state also becomes longer with decreasing resolution, especially for the $\mathcal{R}_i \leq 1$ simulations.

\subsubsection{Over-ionization}
\label{sec:over-ion}
The time lag to reach a steady state is not very important to the subsequent evolution of \HII regions, since most processes we are interested in occur on time-scales far longer than the formation of the initial Str\"omgren sphere. However, the  \textit{over-ionization} issue we mentioned before can significantly increase the total mass of ionized gas in low-resolution galactic and cosmological simulations if we adopt a direct RT feedback implementation.

Equation~(\ref{equ:thermoeq}) regulates the thermochemistry in the pure hydrogen medium, and in the context of UV ionization, the second term on the RHS, corresponding to the collisional process, can be neglected, while the third term can be simplified to a single \HI ionization photon band. The balance equation of the ionization state of the gas can thus be written as
\begin{equation}
    \alpha_\text{B}n_\text{\HII}n_{e}\approx\tilde{c}n_\text{\HI}\sigma_{\text{\HI}}n_{\gamma} \, ,
\end{equation}
where $n_\text{\HII}=1-n_\text{\HI}=n_{e}=x_\text{\HII}n_{\text{H}}$. Substituting the number densities with ionization fractions, we obtain
\begin{equation}
\label{equ:balanceEq}
    \alpha_\text{B} n^2_\text{H}x_\text{\HII}^2\approx\tilde{c}(1-x_\text{\HII})n_\text{H}\sigma_{\text{\HI}}n_{\gamma}  \, ,
\end{equation}
where the LHS corresponds to the recombination rate $R_{\text{rec}}$ and the RHS corresponds to the photoionization rate $R_{\text{ion}}$. 

In high-resolution cases, we suppose that the initial Str\"omgren sphere can be resolved by a series of thin shells. For a cell in the $k$th shell with radius $r_k$, the equilibrium solution of photon density is $n_\gamma = Qe^{-\tau(r_k)}/4\pi r_kc$, where $\tau(r_k)=n_\text{H}\sigma_\text{\HI}\sum_1^kx_\text{\HI}(r_k)\Delta r$ and $\Delta r \sim \Delta x$. With sufficiently high resolution, equation~(\ref{equ:balanceEq}) will approach the analytical solution given by equation~(\ref{equ:OF06}). However, in the low-resolution cases with mass resolution $\mathcal{R}_i<1$, the cell size $\Delta x$ will be larger than $R_\text{S}$. Assuming the entire initial Str\"omgren sphere is embedded in a single cell, the equilibrium balance equation becomes
\begin{equation}
    \alpha_\text{B}n^2_\text{H}x^2_\text{\HII} \approx Q/V_\text{cell} 
\end{equation}
substituting $V_\text{cell}=M_\text{cell}/n_\text{H}m_\text{H}$ with the definition of mass resolution ${\cal R}_i$ (equation~\ref{equ:resolution}), we have
\begin{equation}
    x_\text{\HII} \approx {\cal R}_i^{1/2}.
\end{equation}
Thus, in fully unresolved cases (${\cal R}_i\ll1$) the equilibrium ionization fraction is proportional to ${\cal R}_i^{1/2}$ and the
mass of ionized gas is $\sim M_\text{cell}{\cal R}_i^{1/2}=M_\text{S}{\cal R}_i^{-1/2}$, overestimated by a factor of ${\cal R}_i^{-1/2}$. Therefore, {\it over-ionization} is a direct consequence of recombinations and ionizations attempting to become balanced in a Str\"omgren sphere diluted by the impact of insufficient spatial resolution.

In other words, in low resolution cases, the ionization fraction is supposed to saturate at $x_\text{\HII}\sim\mathcal{R}_i$ to obtain the correct ionized mass $M_S$. However, with this ionization fraction, the ionization rate $R_{\text{ion}}$ is still much larger than the recombination rate $R_{\text{rec}}$ at this time. More gas will be ionized before the $R_{\text{ion}}$ and $R_{\text{rec}}$ are finally balanced at $x_\text{\HII}\sim\mathcal{R}_i^{1/2}$, resulting in the {\it over-ionization}. In section~\ref{sec:Rifix}, we will introduce a method to prevent this issue by enforcing the correct balance between recombination and ionization rates. In section~\ref{sec:where2spatial}, we will discuss the conditions for which this correction is needed.

Similar resolution issues make a significant difference in the expansion phase of the \HII region as well -- we will turn to the early time expansion and discuss these numerical issues in the next section.

\subsection{Test 2 -- Early time expansion and over-heating}
\label{sec:EXPtest}
In the next series of tests, we simulate the expansion of an \HII region in a uniform density medium. The simulation box is initialized with pure hydrogen gas of density $n_{\text{H}}=100$\,cm$^{-3}$ and temperature $T = 100$\,K. A steady source is placed at the centre of the box that emits a blackbody spectrum with $T_{\text{eff}}=35\,000$\,K at a rate of $Q =10^{48}$\,s$^{-1}$, which is similar to a $\sim30\,\Msun$ massive star and the effective \HI ionization cross-section is $3.8\times10^{-18}$\,cm$^2$. We use a reduced speed of light with $\tilde{c}=0.01\,c$ and a multifrequency RT scheme as described in Section~\ref{sec:RT}. We vary the mass resolution $\mathcal{R}_i$ across a range of $0.01$ to $1000$ corresponding to $M_\text{cell}(\Delta x)$ from $4800\,\Msun(12\,\text{pc})$ to $0.048\,\Msun(0.27\,\text{pc})$. To temporally resolve the formation of the initial Str\"omgren sphere, we force the hydrodynamic time-step to be shorter than $0.1\,t_{\text{rec}}$ (0.12\,kyr).

\begin{figure*}
	\includegraphics[width=2\columnwidth]{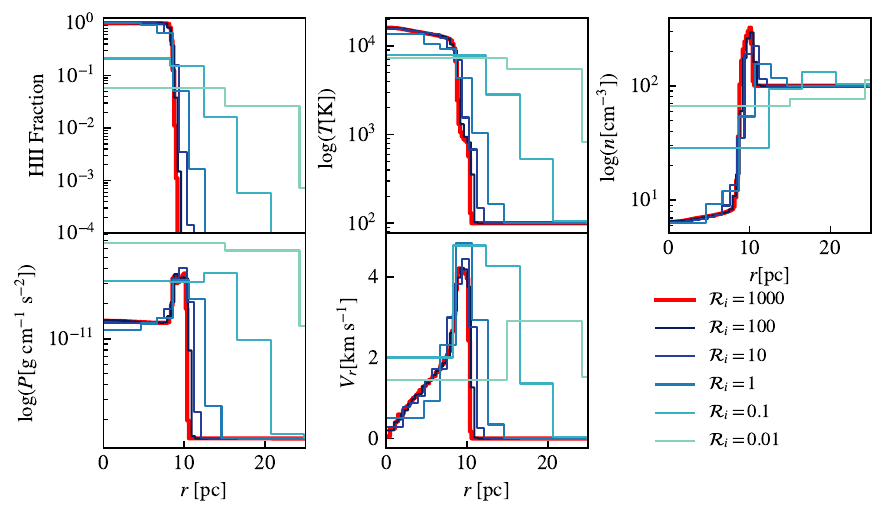}
    \caption{Results of the expansion test (Test~2, Section~\ref{sec:EXPtest}). Shock parameters at 1 Myr: the ionization (top left panel), temperature (top middle panel), density (top right panel), pressure (bottom left panel), and velocity (bottom middle panel) profiles at 1 Myr with different resolutions.}
    \label{fig:test2.1}
\end{figure*}

\begin{figure*}
	\includegraphics[width=2\columnwidth]{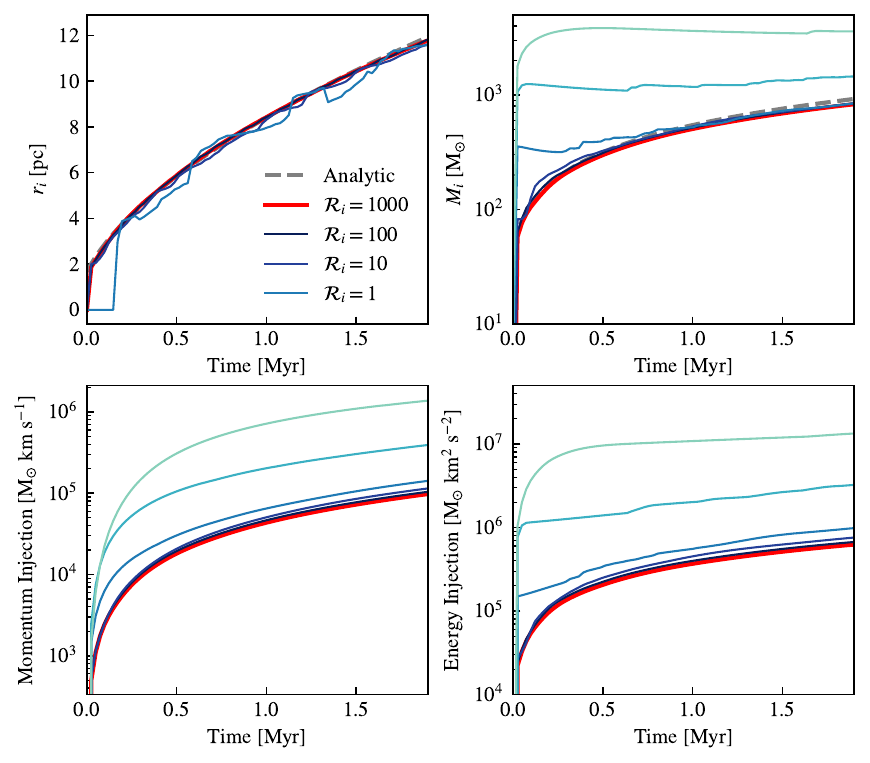}
    \caption{Results of the expansion test (Test~2, Section~\ref{sec:EXPtest}). Evolution of expanding \HII regions: {\it Top left} panel: the Str\"omgren radius as a function of time for simulations with different resolutions; {\it Top right}, {\it bottom left}, and {\it bottom right} panels: the time evolution of ionized mass, momentum injection, and energy injection with different resolutions.}
    \label{fig:test2Summary}
\end{figure*}

In Figure~\ref{fig:test2.1}, we show the profiles of ionization fraction, temperature, density, pressure, and velocity at $1$\,Myr for all the simulations in this test. At this time, the St\"omgren sphere has expanded to $r_i\approx5R_{\rm S}$. In high-resolution simulations like the $\mathcal{R}_i=1000$ run, the profiles show characteristics of a typical D-type front. After the D-type expansion begins, the I-front is always preceded by a hydrodynamical shock front sweeping outward. The interior of the \HII region is evacuated by the expansion and becomes much more rarefied than the surrounding neutral gas. The $\mathcal{R}_i=10$ simulation retains these features with visually acceptable flattening, while lower resolutions gradually flatten these away. In the $\mathcal{R}_i=0.1$ simulation, the gas density is almost uniform, implying that the shock is not captured.

Some delicate features can be captured in the highest-resolution simulation, like a secondary pressure peak behind the shock front and a knee of temperature in the transient region. However, only the $\mathcal{R}_i \geq 100$ simulations can reproduce these secondary structures, as the $\mathcal{R}_i=10$ simulation has already smoothed them out. Nonetheless, the profiles can be roughly reproduced in marginally-resolved cases ($\mathcal{R}_i=10$ and $1$). In simulations conducted at lower resolutions, the initial Str\"omgren sphere is completely unresolved. The insufficient resolution causes all the profiles to appear flatter and results in incomplete ionization mirroring the scenario discussed in Section~\ref{sec:Formation}. Moreover, a substantial fraction of ionized gas also fails to be heated to $10^4$\,K, and the density inside the I-front stays within the same order of magnitude as the background density, and the pressure becomes much higher.

In Figure~\ref{fig:test2Summary}, we present the evolution of the ionization radius, ionized mass, injected momentum, and injected energy. The ionization radius $r_{i}$ (top left panel) is defined as the radius of the shell where $x_\text{\HI}=x_\text{\HII}$ since the profile of the I-front is flattened. The $\mathcal{R}_i\leq1$ runs have ionization fractions that never exceed 0.5 so they are unable to be plotted in this panel. The grey dashed curves in the top-left and top-right panels indicate the Hosokawa-Inutsuka solution (equation~\ref{equ:H-I}). The ionization radius $r_{i}$ matches well with the analytic solution even for the marginally-resolved case ($\mathcal{R}_i=1$), and the ionized mass tends to converge to this solution as the \HII region expands. However, the \textit{over-ionization} issue makes a significant difference in the ionized mass at very early times, as discussed in Section~\ref{sec:Formation}. A surging of the ionized mass can be found in the $\mathcal{R}_i=10$ to $0.01$ results, and it takes an increasing amount of time to relax to the analytic solution with  decreasing $\mathcal{R}_i$. This suggests that the expansion of \HII regions can counteract the overestimation of ionized mass caused by the \textit{over-ionization} to some extent. Even so, the expansion only increases the ionized mass slowly and it can take a considerable time, even longer than the lifetime of the massive star, to counteract the initial surging if the \textit{over-ionization} is too severe.

In the bottom panels of Figure~\ref{fig:test2Summary} we present the cumulative radial momentum and injected energy summed over the entire simulation box. In the low-resolution cases, the momentum and energy injections are also enhanced. Unlike the overestimated ionized mass which tends to asymptote to the analytic solution, the momentum and energy remain consistently larger than the high-resolution solution. 
This is because once momentum and energy are over-injected, no mechanism can eliminate the excess momentum or energy, unlike the ionization state of the gas, which can regulate itself by balancing the photoionization and recombination process.

\subsubsection{Over-heating}
\label{sec:Over-heating}
In this section, we illustrate that the enhancement of momentum and energy injected in low-resolution cases is a result of the artificial heating of a substantial amount of partially ionized gas, raising its temperature to several thousand Kelvin. We refer to this phenomenon as {\it over-heating}. 

The hot gas in the initial Str\"omgren sphere is heated by photoionization and reaches equilibrium with cooling processes at $T\gtrsim10^4$\,K. Then the expansion of the \HII region is driven by the large pressure gradient between the $10^4$\,K hot ionized gas and the surrounding background neutral gas. As the expansion is a hydrodynamic response to the initial ionization, the {\it over-heating} issue is a secondary disaster of the {\it over-ionization} issue we have discussed in Section~\ref{sec:over-ion}.

When the star ignites, the neighbouring gas cells will begin to be ionized and the deposited energy of ionization photons is converted to thermal energy in each step, following equation~\ref{equ:phoheat}. On the other hand, the hot gas will lose internal energy efficiently through various cooling processes. The difference between photoheating and radiative cooling results in a net increase of the internal energy (equation~\ref{equ:DeltaU}). 

In low-resolution cases, the {\it over-ionization} issue leads to an overestimation of the mass of ionized gas but this gas can only be partially ionized because ionization balance cannot be established properly (Section~\ref{sec:Formation}). Similarly, we now discuss the balance between cooling and heating in the initial Str\"omgren sphere. In equilibrium, the heating and cooling rates satisfy
\begin{equation}
   \Gamma = \Lambda = \Lambda_\text{rec}+\Lambda_\text{bb}+\Lambda_\text{bf}+\Lambda_\text{ff} \, ,
\end{equation}
where each rate has the following dependence on the ionization fraction (equations~\ref{equ:H_rec}, \ref{equ:ff}, \ref{equ:H_bb}, and \ref{equ:H_bf}):
\begin{equation}
  \Gamma\propto (1-x_\text{\HII}) \, , \; \Lambda_{\rm rec/ff} \propto x^2_\text{\HII} \, , \; \text{and} \; \Lambda_\text{bb/bf} \propto x_\text{\HII}(1-x_\text{\HII}) \, .
\end{equation}
In low-resolution cases ($\mathcal{R}_i\ll1$), $x_\text{\HII}\sim\mathcal{R}_i$ and the heating and cooling rates can be approximated as follows:
\begin{equation}
  \Gamma\propto 1 \, , \; \Lambda_{\rm rec/ff}\propto \mathcal{R}_i^2 \, , \; \text{and} \; \Lambda_\text{bb/bf}\propto \mathcal{R}_i \, .
\end{equation}
The heating rate stays high in the gas cells even though it should drop down in high-resolution cases where these cells are fully ionized. On the other hand, the cooling rates are all reduced by insufficient ionization, especially those for recombination and Bremsstrahlung.

This scenario bears resemblance to the {\it over-ionization} issue, yet an additional numerical concern exacerbates matters. Besides the ionization fraction, all cooling functions involved are sensitive to the temperature (equations~\ref{equ:H_rec}, \ref{equ:ff}, \ref{equ:H_bb}, and \ref{equ:H_bf}), while the temperature of a gas cell is calculated based on the internal energy $u$ and mean molecular weight $\mu$ by 
\begin{equation}
    T = \frac{2u}{3k_\text{B}\mu m_{\text{H}}}.
\end{equation}
In our pure hydrogen case, the mean molecular weight equals to
\begin{equation}
    \mu = 1/(1 + x_\text{\HII})\sim 1 - \mathcal{R}_i \, , \; \text{when} \, \mathcal{R}_i \ll 1 \, .
\end{equation}
If the initial ionization cannot be resolved, the \textit{over-ionization} will lead to a larger ionized mass but lower ionization fraction, thus higher molecular weight $\mu$ and lower temperature $T$. In the $\mathcal{R}_i\ll1$ unresolved cases the ionization fraction $x_\text{\HII}$ falls from unity to $\sim0$ so that $T$ can be reduced by half for the same internal energy.
For example, in the $\mathcal{R}_i=0.1$ case, the molecular weight is 0.91 and the temperature of the ionized gas only reaches 0.55 of that found in the high-resolution case, assuming the same net energy injection. As we presented in Figure~\ref{fig:test2.1}, in the $\mathcal{R}_i=10$ to $1000$ simulations, the central temperature reaches $\gtrsim1.5\times10^4$\,K, while in unresolved cases it falls to less than $8000$\,K.

The most efficient cooling mechanism in the $10^4$\,K primordial ISM is the bound-bound collision (equations~\ref{equ:H_bb}), which has the largest rate coefficients ($7.5\times10^{-19}$\,erg\,cm$^{-3}$\,s$^{-1}$). However, it is also the most sensitive to the temperature 
\begin{equation}
  \Lambda_\text{bb} \propto \frac{e^{-118348/T}}{1+T^{1/2}_5}
\end{equation}
Therefore, $\Lambda_\text{bb}$ will be drastically reduced if the temperature cannot reach its proper value.
For instance, if the temperature inside the \HII region is $T=8000$\,K, the bound-bound collision cooling rate will be reduced with a factor of $5.5\times10^{-4}$ compared to the $T=16\,000$\,K case.

Consequently, all four primordial cooling mechanisms fail to cool the gas under the combined effect of the above two factors. The total cooling rate is thus drastically underestimated while the heating rate is enhanced. Therefore, the injected thermal energy will be significantly overestimated and leads to an {\it over-heating} problem.

It is crucial to avoid {\it over-ionization} and {\it over-heating} problems in simulations because they both lead to divergent feedback when \HII regions are unresolved, i.e. unresolved \HII regions will provide more feedback rather than less feedback, which hinders the convergence of simulations with different resolutions. In Section~\ref{sec:where2spatial}, we will discuss the specific physical conditions in which {\it over-ionization} and {\it over-heating} problems become particularly relevant.

\subsubsection{Multi-phase gas}
\begin{figure*}
    \includegraphics[width=2\columnwidth]{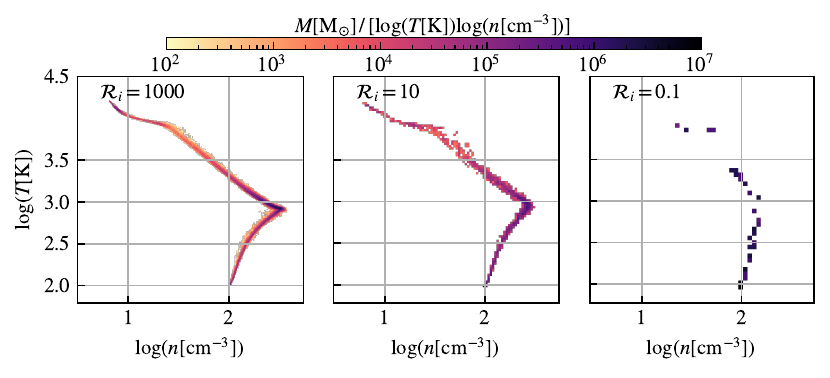}
    \caption{Results of the expansion test (Test~2, Section~\ref{sec:EXPtest}). Mass-weighted phase diagrams for the $\mathcal{R}_i=1000$, $10$, and $0.1$ simulations in Test~2 at $t=1$\,Myr.}
    \label{fig:phase}
\end{figure*}

In Figure~\ref{fig:phase}, 
we present the mass-weighted phase diagram for the $\mathcal{R}_i=1000$, $10$, and $0.1$ simulations at $t=1$\,Myr. The gas shows a continuous distribution over the phase space in the high- and medium-resolution simulations, which separates into several ``islands'' at low resolution. The gas density in the high-resolution case spans a range of two orders of magnitude. The inflection at $\sim10^{2.5}$\,cm$^{-3}$ corresponds to the shock-compressed gas that has accumulated in front of the D-type I-front. Lowering the resolution flattens this inflection, and the gas in the ${\cal R}_i=0.1$ run is distributed close to the vertical line $n_\text{H} = 100$\,cm$^{-3}$ (initial density) with $T<10^4$\,K. This suggests that the shock transition zone and the compressed gas layer between the shock and I-front are not properly resolved.

\begin{figure}
    \includegraphics[width=\columnwidth]{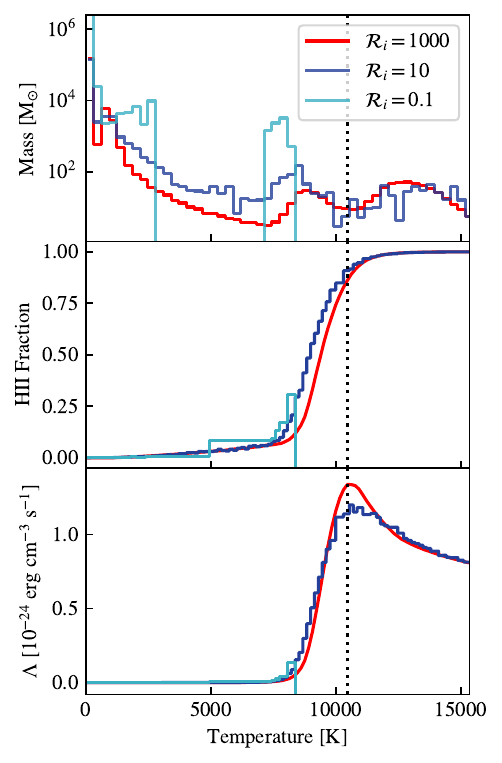}
    \caption{Results of the expansion test (Test~2, Section~\ref{sec:EXPtest}). \textit{Top} panel: mass-weighted histograms of gas mass as a function of temperature for the $\mathcal{R}_i=1000$, $10$, and $0.1$ simulations in Test~2 at $t=1$\,Myr. {Middle} and {bottom} panels: the \HII fraction and cooling rate as functions of temperature obtained from these simulations.}
    \label{fig:TempHist}
\end{figure}

In Figure~\ref{fig:TempHist} {\it top} panel, we present mass-weighted histograms of the gas mass as a function of temperature for the high ($\mathcal{R}_i=1000$), medium (10), and low (0.1) resolution runs at 1\,Myr. As a comparison, we also present the \HII fraction ({\it middle} panel) and cooling rate ({\it bottom} panel) as functions of temperature obtained from these runs.

In the high- and medium-resolution runs, the temperature has a continuous distribution on the $100$\,K to $16\,000$\,K interval of temperature and there are four populations of the different gas phases. The largest population on the left is the undisturbed background gas. The hottest gas populated from about $\sim10\,000$--$15\,000$\,K is the fully ionized gas inside the Str\"omgren sphere. The population at $\sim8000$--$10\,000$\,K is partially ionized and there is a ramp of heated warm neutral gas ranging from $\sim100$--$8000$\,K. These two populations are the shock-heated gas between the I-front and shock front during the D-type expansion, and they are accumulating as a bump around $9000$\,K because of the inefficient cooling. The cooling rate reaches its maximum value at $10\,500$\,K, which results in a valley between the two bumps of populations.

At lower resolutions, the temperature distribution becomes discrete and the valleys are unpopulated. In the $\mathcal{R}_i=0.1$ run, only the $<2500$\,K neutral population and $\sim8000$\,K low ionization population still exist and the $T>10^4$\,K fully ionized population disappears. This gives an intuition for the impact of the {\it over-ionization} and {\it over-heating} issues, as discussed in Section~\ref{sec:Over-heating}. The neutral gas fails to become fully ionized and heat to $>10^4$\,K but stops at $\sim8000$\,K with a low ionization fraction, which is inefficient for cooling. Notice the y-axis is in log-scale so the total mass of ionized gas at low resolution is actually larger than that at high resolution though the \HII fraction is low at that temperature. 

In summary, the insufficient spatial resolution flattens out the shock structures and the {\it over-heating} issue changes the multi-phase gas structure of the ISM by turning the highly-ionized, hot ($>10^4$\,K) gas within \HII regions into partially-ionized, warm ($\sim8000$\,K) gas. Such errors in the gas phase structure can be problematic in line luminosity estimates \citep{2022MNRAS.517....1S,2022MNRAS.513.2904T}. 
Failure to resolve the low-density bubble evacuated by the ionization feedback and the swept-up shell also leads to an inability to reproduce their enhanced or weakened effects on the final momentum output of SN explosions when all these feedback channels are nonlinearly coupled \citep[e.g.][]{2015MNRAS.451.2757W,2016MNRAS.460.2962H}. This leads to uncertainties in the total energy/momentum injection from stellar feedback and the amount of gas blown out of a galaxy.

\section{Effects of different Temporal Resolution}
\label{sec:time-stepping}
Choosing a proper time step is crucial to obtain correct simulation results and minimise computational cost. We now examine the dependence of the simulation results on the choice of the time step. \areport is a moment-based RT implementation treating photons with finite speed of light. The most well-known time-stepping criterion for convergent time integration is the Courant criterion \citep{Courant1928}, and it is modified to involve the reduced speed of light \citep{2019MNRAS.485..117K}
\begin{equation}
    \Delta t_{\rm C} < \eta\,\frac{\Delta x}{\tilde{c}+|{\bm v}_\text{cell}|} \, ,
\end{equation}
where $\Delta x$ and ${\bm v}_\text{cell}$ are the width and velocity (in the lab frame) of a cell and $\eta\sim0.3$ is the Courant factor. In expanding \HII regions, the velocity of a D-type shock is $v_s\lesssim2c_i\ll\tilde{c}$ so
\begin{equation}
    \label{equ:tC}
    \begin{aligned}
    \Delta t_{\rm C} &\approx \eta\frac{\Delta x}{\tilde{c}}=\frac{\eta}{\tilde{c}}\left(\frac{Q}{\alpha_{\rm B} n_\text{H}^2 \mathcal{R}_i}\right)^{1/3} \\
    &\approx230\,\text{yr}\, \left(\frac{\eta}{0.3}\right) \left(\frac{\tilde{c}}{10^{-3}c}\right)^{-1}\left(\frac{{\cal R}_i}{10}\right)^{-1/3}Q_{48}^{1/3}n_3^{-2/3} \, .
    \end{aligned}
\end{equation}
It turns out in practice that such a criterion ensures a stable convergence for hydrodynamic simulations. However, since the mean free path (MFP) of photon ionization is very hard to spatially resolve ($\Delta x<l_\text{mfp}=1/\sigma_\text{\HI}n_\text{H}$), the Courant criterion almost always fails to catch the ionization timescale 
\begin{equation}
    \label{equ:tion}
    \begin{aligned}
        &t_{\rm ion} = \frac{1}{\tilde{c}n_\text{\HI}\sigma_\text{\HI}}=\frac{\alpha_{\rm B}}{\tilde{c}\sigma_\text{\HI}}t_{\rm rec}\, \\
        &\approx1.4\times10^{-3}\left(\frac{\tilde{c}}{10^{-3}c}\right)^{-1}t_{\rm rec}=0.17\,\text{yr}\left(\frac{\tilde{c}}{10^{-3}c}\right)^{-1}n_3^{-1} \, .
    \end{aligned}
\end{equation}

In a forming \HII region, the ionization state of gas is changing drastically, and in a microscopic view, the gas ionization timescale is essentially the $t_\text{ion}$ presented above.
Such a short timescale is difficult to temporally resolve, which can lead to errors in the solutions. We emphasize that this is a general issue across numerical RT  methods and explicit time-stepping schemes though we introduce it from a moment-based scheme with a finite speed of light.

\subsection{Missing photons issue}
In most thermochemistry solvers (including \areport), to reduce the computational cost and ensure numerical stability, the photon transport and the thermochemistry are solved with an operator-splitting strategy \citep[e.g.][]{2009MNRAS.396.1383P,2011MNRAS.414.3458W,2013MNRAS.436.2188R,2019MNRAS.485..117K,Chan2021}.
In the thermochemistry networks, the photon number density $n_{\gamma}$ is not coupled in as a variable but is considered to be a known quantity as the solution of the last photon transport step. The absorption of photons is simply calculated independently of solving the thermochemistry equation with the optical depth in a cell by
\begin{equation}
    \label{equ:photonchange}
        \frac{{\rm d} n_{\gamma}}{{\rm d} t} = -\tilde{c}n_\text{\HI}\sigma_\text{\HI}n_{\gamma} = \frac{n_{\gamma}}{t_\text{ion}} \, ,
\end{equation}
where $n_\text{\HI}$ takes the value at the beginning of the current step.
Once a package of photons is dumped or propagated into a neutral gas cell, the solution for the above equation is simply
\begin{equation}
    \label{equ:photonchangeS1}
    n_{\gamma}(\Delta t) = n_{\gamma,0}\,e^{-\Delta t / t_\text{ion}} \, ,
\end{equation}
where $n_{\gamma,0}$ is the photon density at the beginning of the timestep.

In this paper, we refer to these simplified thermochemistry solvers where photon density is not coupled in the chemistry network as ``uncoupled solvers''.
This simplification has the advantage of efficiency especially when multiple RT bins and transfer of momentum from radiation to gas are implemented and it has been commonly used in various RT implementations.
However, we will show it can lead to a severe numerical error by overestimating the number of absorbed photons, turning them to ``{\it missing photons}''. 

\subsubsection{Missing photons when radiation encounters neutral gas}
\label{sec:missingPhotonInNeutral}

\begin{figure}
	\includegraphics[width=\columnwidth]{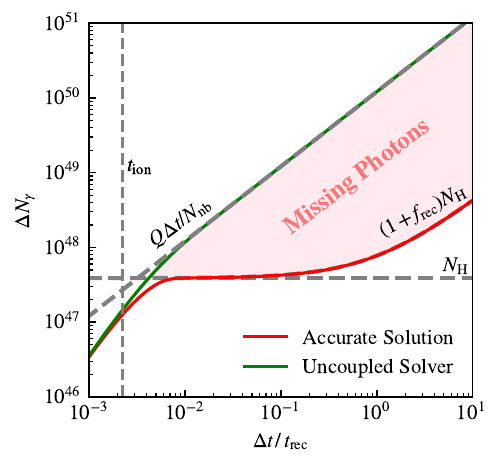}
    \caption{Number of photons absorbed ($\Delta N_\gamma$) in a fully neutral neighbouring cell during the first step after the star ignites, as a function of the time step size $\Delta t$ (assuming the sizes of all the relative time steps match each other). The red and green curves are the accurate solution obtained with the coupled solver (see Section~\ref{sec:CouplingphotonDensity}) and the problematic solution obtained with the uncoupled solver. The pink-filled region presents the {\it missing photons} described by equation~(\ref{equ:dNg}).}
    \label{fig:MissingPhotonLines}
\end{figure}

When the star ignites, ionizing photons are dumped into the predominantly neutral gas and propagate outwards. We use a one-zone case to demonstrate the {\it missing photons} issue when photons are absorbed in a fully neutral neighbouring cell of $n_\text{H}=10^9$\,cm$^{-3}$ during the first step after a $Q=10^{48}$\,s$^{-1}$ star ignites. In Figure~\ref{fig:MissingPhotonLines}, we plot the number of photons absorbed ($\Delta N_\gamma$) in a fully neutral neighbouring cell during the first step after the star ignites, as a function of the time step size $\Delta t$ (assuming the sizes of all the relative time steps match each other). The red curve is obtained by an accurate thermochemical solver solving the photon and ion numbers simultaneously (see Section~\ref{sec:CouplingphotonDensity}), we regard this solution as the accurate one for our question. On the other hand, the green curve is obtained by the widely used uncoupled solver.

In each step when radiation encounters neutral gas, the number of photons absorbed in this step is (equation~\ref{equ:photonchangeS1})
\begin{equation}
    \Delta N_{\gamma} = N_{\gamma}(1-e^{-\Delta t/t_{\rm ion}})\approx N_{\gamma}\, ,
\end{equation}
where $N_{\gamma}$ is the number of photons prior to absorption, $\Delta t/t_{\rm ion}\gg1$ in most cases so the second equality holds. This equation demonstrates that almost all photons dumped or propagated into the gas will be consumed when the $t_\text{ion}$ (or MFP) is unresolved. As shown by the green curve in Figure~\ref{fig:MissingPhotonLines}, the number of absorbed photons converges to the number of injected photons $Q\Delta t/N_\text{nb}$ when the time step size is larger than several $t_\text{ion}$. 

On the other hand, the total number of neutral hydrogen atoms in a cell is
\begin{equation}
    N_\text{\HI} = \frac{M_{\rm cell}}{m_{\rm H}}=\frac{Q}{\alpha_{\rm B}n_{\rm H}{\cal R}_i} \, .
\end{equation}
Assuming one photon can ionize one neutral hydrogen atom (MFP unresolved), in this case, if $\Delta N_{\gamma}>(1+f_\text{rec})N_\text{\HI}$, then $\Delta N_{\gamma}$ photons will be absorbed but only $N_\text{\HI}$ atoms can be ionized, the photon number is no longer conserved. Here $f_\text{rec}$ is a factor accounting for the recombination of ions. In other words, $\delta N_{\gamma}$ photons simply disappear without any effect on the gas state, they become the {\it missing photons} (pink-filled region in Figure~\ref{fig:MissingPhotonLines}). Specifically, the number of missing photons is
\begin{equation}
    \label{equ:dNg}
    \delta N_{\gamma} = \Delta N_{\gamma}-(1+f_\text{rec})N_\text{\HI} \, ,
\end{equation}
where $f_\text{rec}$ can be estimated as 
$\Delta t/t_\text{rec}$ because $\Delta t\sim t_\text{rec}\gg t_\text{ion}$.
If the size of the time step is too large, $\Delta N_{\gamma}\propto Q\Delta t\gg(1+f_\text{rec})N_\text{\HI}$, almost all the photons are lost, having no effect on the ionization state when they enter a neutral gas cell.

In Appendix~\ref{sec:step-by-step}, we track the step-by-step events for the {\it missing photons} issue by considering an idealized scenario (Figure~\ref{fig:missingphoton}). We find that the following criterion should be satisfied to avoid the occurrence of {\it missing photons} if the photon density is not coupled in the thermochemistry solver:
\begin{equation}
\label{equ:Kcrit}
        \Delta t \lesssim  t_\text{rec} \frac{N_{\rm nb}}{{\cal R}_i} \qquad \left(\text{or} \quad {\cal R}_i \lesssim N_{\rm nb} \quad \text{if} \quad \Delta t \gtrsim t_\text{rec}\right) \, .
\end{equation}
This criterion is stringent because it suggests the mass resolution should not be higher than $N_\text{nb}$ or the size of the time step should be far smaller than $t_\text{rec}$.
Nonetheless, the {\it missing photons} issue may not be as troublesome as long as the photon injection and thermochemistry have the same cadence. This only occurs when a large number of photons are being dumped or propagated into a substantially neutral cell.
However, realistic implementations usually conduct photon injection, radiative transport, and thermochemistry with different cadences. In the next section, we will show that the {\it missing photons} issue can lead to disastrous consequences in these cases.

\subsubsection{Mismatching cadences make things worse}
\label{sec:Subcycling}

\begin{figure}
	\includegraphics[width=\columnwidth]{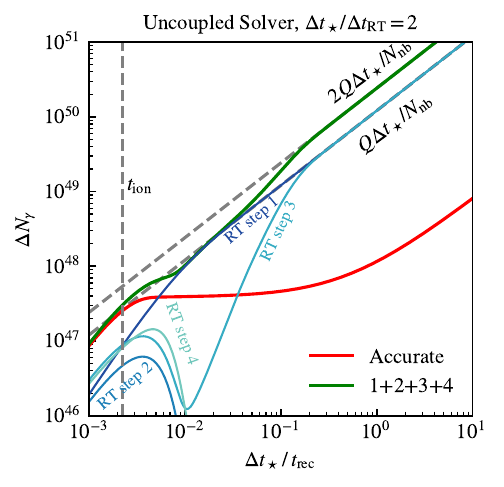}
    \caption{Similar to Figure~\ref{fig:MissingPhotonLines}, but assume $\Delta t_\star=2\Delta t_\text{RT}$ and evolve it to the end of the second injection step after the star ignites. The red and green curves are the cumulative number of photons absorbed in these two inject steps (four RT steps) obtained by the accurate and uncoupled solver. The slightly thinner curves are the number of photons absorbed in the four RT steps with the uncoupled solver, the summation of these four curves is the green line.}
    \label{fig:UncouplAnalytic}
\end{figure}

In realistic simulations, it is common to encounter mismatching cadences between photon injection, radiative transport, and thermochemistry. Typically, the frequency of photon injection is determined by the star particle time step, while the RT and thermochemistry time steps of the neighbouring cells can be further refined. 
Consequently, photon dumping is typically performed with a relatively low cadence compared to RT, i.e., each photon injection step can be associated with several RT steps. Further subcycling for thermochemical integration is also common when the gas properties change drastically in a single RT step, leading to a more severe mismatch between photon injection and thermochemistry cadences.

To illustrate the problem, we consider the case of multiple subcycles of RT and thermochemistry conducted within one injection time step \citep[as described in][]{2019MNRAS.485..117K}.
In our tests, we use an equal time-stepping scheme, so the injection time step of the star $\Delta t_\star$ is equal to the hydrodynamic time step $\Delta t$, and the subcycled RT step $\Delta t_\text{RT} = \Delta t/N_\text{sub}$. Even if one does not intentionally subcycle the RT, the two-stage Heun's integration method introduces at least two ``substeps'' for RT, making this a potential issue for the similar variant of second-order implementations, regardless of whether they adopt explicit RT subcycling or not.

In Figure~\ref{fig:UncouplAnalytic}, we present a similar one-zone test as the one in the last section but now assuming $\Delta t_\star = 2\Delta t_\text{RT}$. We evolve the system to the end of the second injection step after the star ignites. It is important to note that the {\it missing photons} issue now occurs not only in the first RT step (`RT step 1') immediately after the first injection of photons but also in subsequent injection steps. The difference is that fresh photons are not injected immediately after this RT step. However, in the remaining second RT step (`RT step 2'), the photon density $n_{\gamma}$ is almost $0$ if the {\it missing photons} issue has occurred. As a result, ionization processes become negligible and the gas begins to recombine. The \HII fraction before the next injection step $x_\text{inj}$ can be estimated as
\begin{equation}
    \label{equ:xprime}
     x_\text{inj} = \frac{1}{1+\Delta t_\text{RT}/t_\text{rec}} \, .
\end{equation}

Consequently, the gas will not be fully ionized even after the occurrence of the {\it missing photons} issue. The partially neutral gas will keep eliminating photons that enter this cell in subsequent time steps. Similar to equation~(\ref{equ:dNg}), the number of {\it missing photons} in the first substep of the next sourcing step (denoted as step~2) is
\begin{align}
\label{equ:dN2}
    \delta N_{\gamma,2} = N_{\gamma,2}\left[1-e^{-(1-x_\text{inj})\Delta t_\text{RT}/t_\text{ion}}\right]\notag
    \\-\left[1-x_\text{inj}+f_\text{rec}(\Delta t_\text{RT})\right]N_\text{\HI} \, ,
\end{align}
where $N_{\gamma,2}$ is the number of photons in the second step before absorption. 

As we present in Figure~\ref{fig:UncouplAnalytic} with the `RT step 3' curve, the second injection step will meet a severe {\it missing photons} issue if $\Delta t_\star\gtrsim0.1t_\text{rec}$. This leads to all $2Q\Delta t_\star$ photons injected in these two steps being absorbed after the third RT step (`RT step 3'), leaving almost no photons in the gas.
If the number of survived photons is insufficient to suppress the recombination in the fourth RT step (`RT step 4'), the \HII fraction will decrease again as equation~(\ref{equ:xprime}). The {\it missing photons} issue has now led to an instability: the \HII fraction oscillates between 1 and $x_\text{inj}$ (equation~\ref{equ:xprime}) with a period of two RT steps and the photon density does not increase after the first occurrence of {\it missing photons} but remains near zero. In the most severe case (e.g. the $t_\star/t_\text{rec}\gtrsim0.2$ part of Figure~\ref{fig:UncouplAnalytic}), all photons injected subsequently suffer the same severe {\it missing photons} issues caused by the unstable \HII fraction as that in the first two injection steps. 
Even though a fraction of photons may escape, they potentially also meet the same issue in subsequent neutral cells. Consequently, a substantial amount of photons emitted by the star can be dissipated (see Figure~\ref{fig:criterion}) and lead to an incorrect \HII region and significantly reduced feedback.

Recall that $\Delta t_\text{C}$ is typically $10^3$ times larger than $t_\text{ion}$ (equations~\ref{equ:tC}~and~\ref{equ:tion}), therefore even a relatively high but not fully ionized \HII fraction can consume a large fraction of photons in the first RT step. In Figure~\ref{fig:criterion}, we present the number of surviving photons as a function of $\Delta t_\text{RT}$, where the survival ratio decreases rapidly as $\Delta t_\text{RT}$ increases.

Generalizing equation~(\ref{equ:xprime}), if we have $N_\text{sub} > 1$ but only inject photons once at the beginning, the \HII fraction after subcycling is $x_\text{inj}=1/[1+(N_\text{sub}-1)\Delta t_\text{RT}/t_\text{rec}]$. Thus, we obtain a time-stepping criterion $\Delta t_\star\lesssim0.1t_\text{rec}N_\text{sub}/(N_\text{sub}-1)$, which can ensure $x_\text{inj}>0.9$ to avoid oscillations in the \HII fraction and repeated {\it missing photons}. We validate this criterion quantitatively in Appendix~\ref{sec:validate}, and we also show this leads to converged results in Section~\ref{sec:withoutsubcycling}. This also suggests that having a different cadence between photon injection and thermochemistry may lead to serious consequences, but increasing the number of subcycles should not alter the behaviour significantly.

\subsection{Test 3 -- Consequences of repeating missing photons issue}
\label{sec:withoutsubcycling}
In this experiment, we simulate the formation and expansion of an \HII region in the uniform medium with a density of $10^9\,\text{cm}^{-3}$. Such a high density is set to break the Courant criterion and it has been observed in hyper compact \HII regions \citep[e.g.][]{2021A&A...650A.142M}. The simulation box is initialized with pure
hydrogen gas of temperature $T=100$\,K and the radiation source is the same as that used in Section~\ref{sec:EXPtest} with $T_{\rm} =35\,000$\,K and $Q=10^{48}$\,s$^{-1}$. The reduced speed of light is set as $0.001c$ for a larger $\Delta t_{\rm ion}$. The mass resolution $\mathcal{R}_i$ is set as $10^4$ to exclude the effects of low spatial resolution. We minimize the number of RT subcycles as $N_\text{sub}=2$, which is an intrinsic requirement of the second-order scheme of \arepo. We use a forced equal time-stepping scheme so that the RT time steps for all the cells are the same throughout the simulation and equal to the system integration time step. 

With this simulation setup, we will present the reduced feedback issue when the repeating {\it missing photons} issue happens and test the maximum available time-step to obtain correct momentum feedback with an uncoupled solver.
The recombination time is $t_\text{rec}=1.22\times10^{-4}$ \,yr such that the ionization and Courant times are $\{t_\text{ion}, t_\text{C}\}=\{2.23\times10^{-3},18.9\eta\}\,t_\text{rec}=\{2.77\times10^{-7},2.31\times10^{-3}\eta\}$\,yr.

\label{sec:MaxTimeStep}

\begin{figure*}
	\includegraphics[width=2\columnwidth]{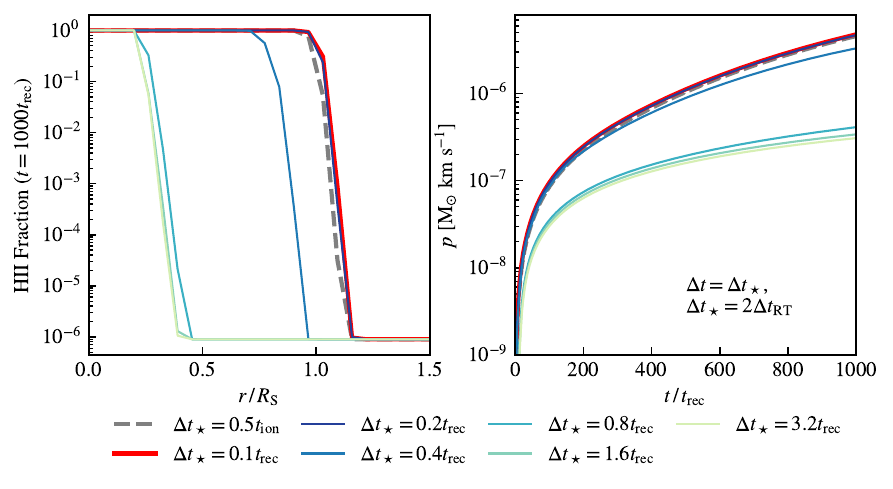}
    \caption{Results of the temporal resolution test (Test~3, Section~\ref{sec:withoutsubcycling}). The {\it left} panel shows the radial profiles of the \HII fraction for the $\Delta t_\star = 0.5\,t_{\rm ion}$ (fiducial) run and $\Delta t_\star = \{0.1,0.2,0.4,0.8,1.6,3.2\}\,t_{\rm rec}$ runs, and the {\it right} panel shows the evolution of injected momentum over $1000\,t_{\rm rec}$.}
    \label{fig:TimestepFig1}
\end{figure*}
To determine the maximum available time step for simulating the feedback from the \HII region, we run simulations with different time steps $\Delta t$ to investigate the effect of temporal resolution. We focus on the radial momentum injection as a quantitative way to evaluate the impact on feedback.

In the left panels of Figure~\ref{fig:TimestepFig1}, we present radial profiles of the \HII fraction and temperature from the $\Delta t_\star = 0.5\,t_{\rm ion}$ (fiducial) and $\Delta t_\star = \{0.1,0.2,0.4,0.8,1.6,3.2\}\,t_{\rm rec}$ runs at $t = 1000\,t_{\rm rec}$. In the right panel, we show the evolution of the cumulative injected momentum over $1000\,t_{\rm rec}$. We find that only runs with a time-step of $\Delta t_\star \leq 0.2\,t_{\rm rec}$ can correctly reproduce the fiducial result. When increasing the time-step to $0.4\,t_{\rm rec}$, the Str\"omgren sphere becomes smaller and the momentum injection is reduced by a factor of $1.5$. With larger $\Delta t_\star$, the size of the \HII region and momentum injection suddenly plunge to below half of the fiducial solution. This is because the $0.8\,t_{\rm rec}$ and larger $\Delta t$ runs fall due to the severe repetition of the {\it missing photons} issue, as presented in Section~\ref{sec:Subcycling}.

In summary, we find that $\Delta t_\star \lesssim 0.1\,t_\text{rec}N_\text{sub}/(N_\text{sub}-1)$ could be a useful criterion for the RT time-step to obtain the correct \HII region feedback under the current implementation of \areport{}. A larger RT time step will result in too little momentum injection, ionization, and heating. 

We note that while our ideal experiment can present convergent results with a time step that satisfies this criterion, in real simulations, gas cells are also allowed to have different time steps. This creates the possibility of photons flowing from a cell with a longer time step into one with a shorter time step, resulting in the occurrence of recurring {\it missing photon} issues.  Therefore, it is crucial to devise a method to correct this issue. In section~\ref{sec:Temporalcorrection}, we will discuss how to deal with this difficulty with the photon density coupled thermochemistry solver and a corrected uncoupled solver.

\section{Correcting for inadequate resolution}
In this section, we will introduce several methods to correct these spatial and temporal resolution problems.
\label{sec:corrections}
\subsection{Spatial correction}
\subsubsection{Lowering number of neighbours during photon injection}
\label{sec:lowerNnb}
In Section~\ref{sec:resolution}, we have found that ${\cal R}_i\gtrsim10$ is needed to ensure the correct ionized mass and feedback while lower resolution leads to increased ionized mass and enhanced feedback. The simplest way to correct the resolution issues is to reduce the number of neighbouring cells that photons from the star are injected into. 

In the photon injection routine, $\Delta N_{\gamma}$ photons from the star particle are dumped into a given number of its nearest
neighbouring gas cells in a weighted fashion. In our default setup, it is set to 32, in order to enshroud the stellar particle.
If one does not care about the morphology of individual \HII regions but only their cumulative feedback to the ISM and galaxy, the spatial resolution problem can be partly solved by lowering the value of $N_{\rm nb}$ to $N_{\rm nb}<{\cal R}_i$. In practice, $N_{\rm nb}$ can be simply set as $2$ to maximize the probability of resolving the Str\"omgren spheres \citep[e.g.][]{2020MNRAS.499.5732K}.

However, this correction cannot eliminate the possibility of {\it over-ionization} so that highly unresolved \HII regions will still lead to divergent feedback. Also, this correction is a global method which is hard to adjust dynamically based on the local resolution. Therefore, we will introduce a novel correction method in the next section, which corrects the resolution effects and provides convergent results by dynamically reducing the ionization rate in the neighbouring cells.

\subsubsection{Accurate ionization and heating balances}
\label{sec:Rifix}
In this section, we describe an alternative way to correct spatial resolution issues based on enforcing the correct balance between recombination and ionization rates (equation~\ref{equ:balanceEq}). 

Recall in Section~\ref{sec:resolution}, when we discussed the resolution effects in low-resolution cases, we assumed that the entire initial Str\"omgren sphere is embedded in a single cell. In this case, when the ionized mass reaches the Str\"omgren mass $M_\text{S}$, the ionization fraction in this cell only reaches ${\cal R}_i$ rather than $\sim1$, which leads to an ionization rate $R_\text{ion}$ larger than the recombination rate $R_\text{rec}$. Our modification aims to reduce the ionization rate by a factor $f_\text{cor}$ so that the ionization and recombination are balanced once the ionized mass reaches $M_\text{S}$, i.e.
\begin{equation}
\label{equ:corrected_balance}
    \alpha_\text{B} n^2_\text{H}x_\text{\HII}^2=f_\text{cor}\tilde{c}(1-x_\text{\HII})n_\text{H}\sigma_{\text{\HI}}n_{\gamma}  \, ,
\end{equation}
when $M_i = M_\text{S}$.

In practice, photons from the star are dumped into $N_\text{nb}$ cells with a weight factor $w_k$ for each neighbour, e.g. based on solid angle weighting. Here we assume that photons from the star are always trapped in the nearest neighbours of the star when the resolution is lower than ${\cal R}_i=N_\text{nb}$. 
In this case, each neighbouring cell $k$ can thus be regarded as an individual unresolved \HII region with a source rate of $Q_k = w_k Q$. By definition of the mass resolution (equation~\ref{equ:resolution}), we have ${\cal R}_{i,k}$ for cell $k$ of
\begin{equation}
\label{equ:Rilocal}
    {\cal R}_{i,k}  = \frac{m_\text{H}}{\alpha_\text{B}}\frac{w_kQ}{n_\text{H} M_\text{cell}} = w_k {\cal R}_i \, .
\end{equation}
When the initial Str\"omgren sphere is formed, the ionized gas of mass $M_i$ is hosted by these $N_\text{nb}$ cells, and $M_i = \sum_k x_{\text{\HII},k} M_\text{cell}$ where $x_{\text{\HII},k}$ is the ionization fraction of cell $k$. Because $\sum_k w_k = 1$, we have $x_{\text{\HII},k}\approx{\cal R}_{i,k}$ at the time when $M_i = M_\text{S}$. This is similar to the idealized one-zone case we discussed in Section~\ref{sec:resolution}, except that we use ${\cal R}_{i,k}$ to account for every neighbour. Replacing $x_{\text{\HII}}$ with ${\cal R}_{i,k}$ in equation~(\ref{equ:corrected_balance}), we have the target balance equation for each cell 
\begin{equation}
    \alpha_\text{B} n_\text{H}{\cal R}_{i,k}^2\approx f_\text{cor}\tilde{c}\sigma_{\text{\HI}}n_{\gamma} \, .
\end{equation}
Therefore, the correction factor $f_\text{cor}$ that is required to establish an accurate balance once $M_i$ reaches $M_\text{S}$ is
\begin{equation}
\label{equ:f_cor}
    f_\text{cor} = \frac{\alpha_\text{B} n_\text{H}{\cal R}_{i,k}^2}{\tilde{c}\sigma_{\text{\HI}}n_{\gamma}} \, .
\end{equation}
The steps to implement this correction are listed below:
\begin{enumerate}
\item We use an attribute ${\cal R}_{i,l}$ to store the local mass resolution of each gas cell $l$ and initialize it as ${\cal R}_{i,l}=-1$ at the beginning of each step. 
\item When looping through neighbouring gas cells around the star in the photon injection routine, evaluate the local mass resolution ${\cal R}_{i,k}$ for each cell $k$ based on its weight factor $w_k$ and gas attributes (equation~\ref{equ:Rilocal}). 
\item In the thermochemistry routine, for each cell $l$, we check whether $0<{\cal R}_{i,l}<1$. If so, we go to the next step to conduct the resolution correction. Otherwise, the cell is either already resolved (${\cal R}_{i,l}>1$) or not a neighbouring cell (${\cal R}_{i,l}=-1$).
\item If cell $l$ requires a resolution correction, activate the correction by checking if $x_{\text{\HII},l}>{\cal R}_{i,l}$, calculate the correction factor $f_\text{cor}$ (equation~\ref{equ:f_cor}), and reduce the ionization and heating rates by multiplying them with $f_\text{cor}$.
\end{enumerate}

In this way, the spatial resolution correction is implemented dynamically only for cells in the neighbour lists of the stars and only based on their own gas attributes. This correction algorithm can thus be easily applied to realistic simulations with non-uniform gas density. The analogous correction factors for more species can also be obtained by modifying the detailed balance equations. We derive and describe the correction factors for He and \ce{H2} in Appendix~\ref{sec:fcor}.

\begin{figure}
	\includegraphics[width=\columnwidth]{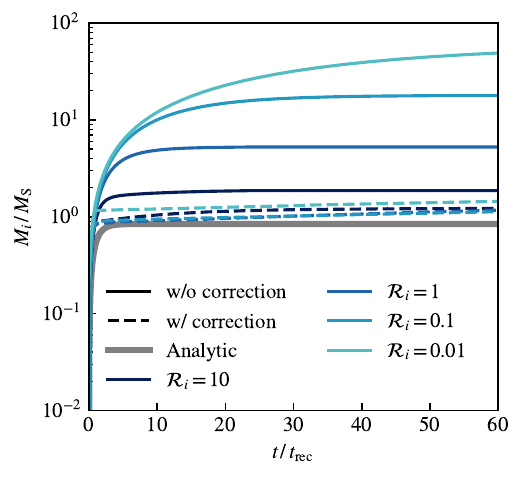}
    \caption{Evolution of the ionized mass from low mass resolution static St\"omgren sphere simulations (Test~1, Section~\ref{sec:Formation}) with (dash curves) and without (solid curves) our correction}
    \label{fig:RiFixT1}
\end{figure}

In Figure~\ref{fig:RiFixT1}, we present the low resolution runs in the static Str\"omegren sphere test (Test 1, Section~\ref{sec:Formation}) with (dash curves) and without (solid curves) our correction. With the correction, both the marginally resolved (${\cal R}_{i}=1$ and $100$) and unresolved (${\cal R}_{i}=0.1$ and $0.01$) Str\"omegren spheres can obtain equilibrium ionized masses within a factor of $2$ of the expected analytic result. We only conduct resolution corrections for cells in the neighbour list of the star, the ionized mass is increasing over time slowly because a portion of photons leaks out of these cells. However, we will soon see that this only leads to a minor effect on the overall momentum feedback.

\begin{figure}
    \includegraphics[width=\columnwidth]{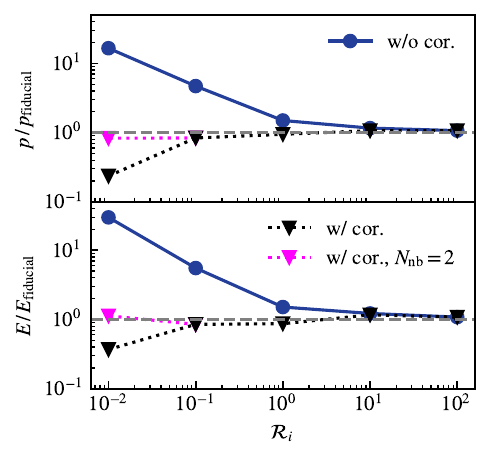}
    \caption{The momentum ({\it top}) and energy ({\it bottom}) feedback at $1$\,Myr obtained with (black triangles) and without (blue circles) the spatial resolution correction as a function of spatial resolution $\mathcal{R}_i$. The magenta triangle shows the results of the ${\cal R}_i=0.01$ run combining reducing the number of neighbours to 2 and our spatial resolution correction. The y-axis quantities are normalized to the fiducial results (${\cal R}_i = 1000$).}
    \label{fig:momFrac}
\end{figure}

In Figure~\ref{fig:momFrac}, we present the momentum and energy feedback with and without the spatial resolution correction at $1$\,Myr.
For the marginally-resolved cases ($0.1 \lesssim {\cal R}_i \lesssim 10$), the momentum and energy feedback after correction perfectly match the high resolution results.
In the case of significant unresolved simulations (${\cal R}_i=0.01$), the inadequate mass resolution results in an artificial enhancement of momentum and energy feedback by over one order of magnitude arising from the effects of {\it over-ionization} and {\it over-heating}.
With the resolution correction, the momentum and energy feedback are much under control and are even reduced by a factor of a few compared to the true solution. This is because our correction not only ensures the correct initial Str\"omgren mass but also partly fixes the {\it over-heating} issue by limiting the photoheating rate and correcting the balance between heating and cooling so the internal energy is approximately conserved (this is not guaranteed because the reduced cooling rate is not corrected). In return, the unresolved Str\"omgren spheres reach even lower temperatures, several hundred Kelvin for the ${\cal R}_i=0.01$ run, leading to insufficient pressure gradients to drive the expansion by converting thermal energy to kinetic energy. Therefore, the highly unresolved \HII regions simply dissolve into the background ISM and provide less amount of feedback. 

To maximize the possibility of resolving ionization feedback, we reduce $N_\text{nb}$ to 2 (Section~\ref{sec:lowerNnb}) along with our resolution correction (Section~\ref{sec:Rifix}). The magenta triangle in Figure~\ref{fig:momFrac} shows the corrected results of the ${\cal R}_i=0.01$ run combining these two methods. 
By concentrating the deposition of ionizing photons in fewer cells, we achieve more accurate momentum feedback and energy injection compared to only using the correction.

\subsubsection{Caveat: Trade-off between correcting ionization feedback and achieving accurate recombination line emission}

Although enforcing the correct balance between recombination and ionization rate can help us obtain the correct ionized mass and feedback from unresolved \HII regions, this can lead to additional issues in the post-processing prediction of recombination line emission, as it preserves the conversion of ionizing photons into line photons \citep[e.g.][]{2022MNRAS.517....1S}. 

Without our correction, the balance between the total ionizing flux and integral recombination rate
\begin{equation}
    \label{equ:Rrec}
    Q = \int R_\text{rec}{\rm d}V = 4\pi \int\alpha_\text{\HII}n_\text{H}^2x_\text{\HII}^2r^2{\rm d}r\,,
\end{equation}
which holds for all resolutions because the photon number is conserved. However our correction equivalently reduces $Q$ with a factor of ${\cal R}_{i,k}$ to ensure
\begin{equation}
    M_i = 4\pi m_\text{H} \int n_\text{H}x_\text{\HII}r^2{\rm d}r = M_\text{S}\,.
\end{equation}
Thus, there is a trade-off between obtaining the correct ionized momentum/energy on the fly based on the correct $M_i$ and obtaining precise recombination line emission with post-processing based on the recombination integral (equation~\ref{equ:Rrec}). 

\begin{figure}
	\includegraphics[width=\columnwidth]{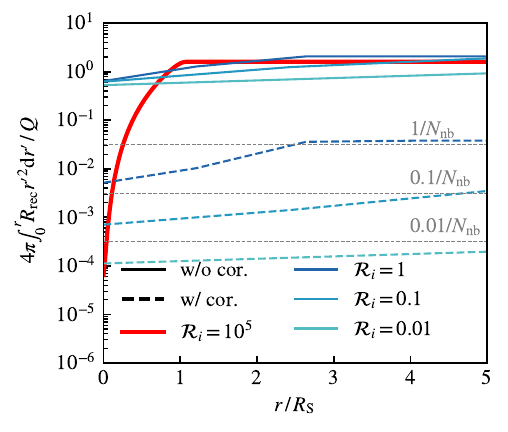}
    \caption{Radial integration of the recombination rate (RHS of equation~\ref{equ:Rrec}, normalized to the ionizing photon rate $Q$) as a function of radius obtained from the runs presented in Figure~\ref{fig:RiFixT1}. The grey dashed lines indicate the value of ${\cal R}_{i,k}$, to which the corrected runs asymptote.}
    \label{fig:Rrec}
\end{figure}
In Figure~\ref{fig:Rrec}, we present the radial integration of recombination rate as the function of radius obtained from the runs presented in Figure~\ref{fig:RiFixT1}. The recombination integrals in the uncorrected runs converge to unity at all the resolutions while those in the corrected runs decrease with a scaling $\propto{\cal R}_{i,k}$. Consequently, employing our correction will lead to an underestimate in line emissivity with a factor of ${\cal R}_{i,k}$. Correcting ionization feedback while simultaneously achieving accurate recombination line emission presents a challenging trade-off. A possible solution is to use the ${\cal R}_{i,k}$ values stored in the gas cells to further correct the recombination rate in the cells where the spatial resolution correction is activated. Ongoing investigations aim to improve solutions to this dilemma.

\subsection{Temporal correction}
\label{sec:Temporalcorrection}

\subsubsection{Coupling photon density into thermochemistry calculations}
\label{sec:CouplingphotonDensity}
The most direct (but expensive) solution to solve the temporal resolution problems is coupling the photon density into the ODE system of thermochemistry calculations. For hydrogen-only cases with a single UV bin, there are three variables: the $\HII$ fraction $x_\text{\HII}$, the photon density normalized to its initial value
$r_\gamma = n_{\gamma}/n_{\gamma,0}$, and specific internal energy $u$. Thus, we need to solve the following three equations simultaneously
\begin{align}
    \frac{\mathrm{d}x_\text{\HII}}{\mathrm{d}t} &=-\alpha_\text{\HII}n_\text{H}x^2_\text{\HII}+\sigma_{\text{e\HI}}n_\text{H}x_\text{\HI}x_\text{\HII} + \tilde{c}\sigma_{\text{\HI}}x_\text{\HI}n_{\gamma,0}r_\gamma \, ,  \notag \\
    \frac{\mathrm{d}r_\gamma}{\mathrm{d}t} &= -\tilde{c}n_\text{H}\sigma_{\text{\HI}}x_\text{\HI} \, , \quad \text{and} \\
    \frac{\mathrm{d}u}{\mathrm{d}t}&=\frac{1}{\rho}\Bigr[\bar{\Gamma}x_\text{\HI}r_{\gamma}-(\Bar{\Lambda}_\text{ff}+\Bar{\Lambda}_\text{rec})x^2_\text{\HII} - (\Bar{\Lambda}_\text{bb}+\Bar{\Lambda}_\text{bf})x_\text{\HI}x_\text{\HII} \Bigr] \,\notag ,
\end{align}
where the $\bar{\Gamma}$ is the constant part of the photoionization heating rate and 
$\bar{\Lambda}$ is the temperature-dependent part of the heating and cooling rates listed in Appendix~\ref{sec:cooling}, respectively.

\begin{figure}
	\includegraphics[width=\columnwidth]{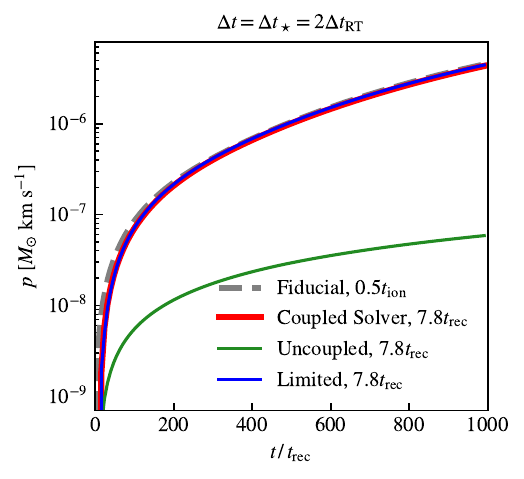}
    \caption{Evolution of momentum injection for Test~3 (Section~\ref{sec:withoutsubcycling}) runs obtained with the coupled solver (red), the uncoupled solver without correction (green), and the uncoupled solver with limited absorption correction (blue), using a Courant time step decided by the code. The grey dashed curve is the fiducial result obtained with a time step $\Delta t = 0.5\,t_\text{ion}$.}
    \label{fig:CoupledSolver}
\end{figure}

In principle, the photon density coupled thermochemistry solver should give an accurate solution no matter how long the integration step is. In Figure~\ref{fig:CoupledSolver}, we present the result of Test~3 (Appendix~\ref{sec:withoutsubcycling}) solved by a coupled solver (red curve). The integration time step is decided by the code as $\Delta t = 7.8\,t_\text{rec}$. This solution matches the $\Delta t = 0.5\,t_\text{ion}$ fiducial run (grey dash curve) perfectly except at the very beginning, where the too-long time step leads to a delay in the formation of the initial Str\"omgren sphere. As a comparison, the green curve shows the solution of the uncoupled solver with the same time step, whose momentum feedback is erroneously reduced by over two orders of magnitude.

\subsubsection{Limited photon absorption in an uncoupled solver}
\begin{figure*}
	\includegraphics[width=2\columnwidth]{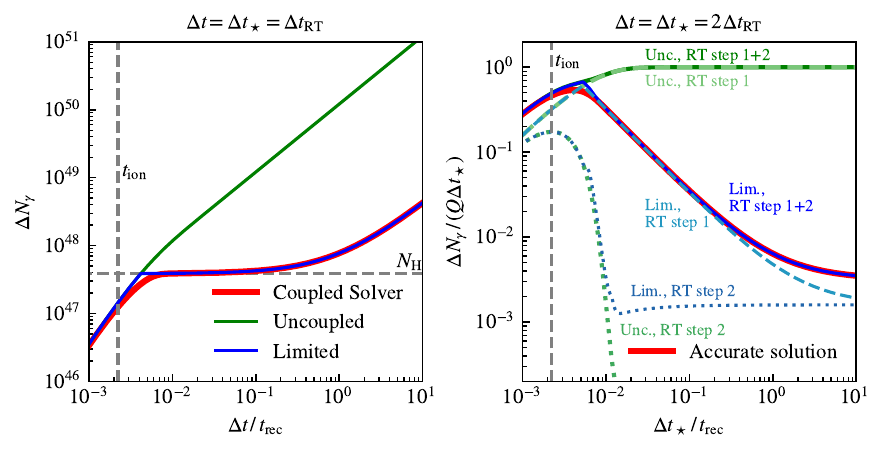}
    \caption{{\it Left panel}: Similar to Figure~\ref{fig:MissingPhotonLines}, but including results from the uncoupled solver with the limited absorption correction (blue curve), which is in agreement with the accurate coupled solver (red curve) avoiding the {\it missing photons} issue seen in the uncoupled solver without correction (green curve).
    {\it Right panel}: Similar to the {\it left} panel but $\Delta t=\Delta t_\star = \Delta t_\text{RT}$ and the y-axis is normalized to the number of photons dumped in an injection step with size $\Delta t_\star$. The red curve shows the accurate solution of the coupled solver and the blue (green) dash, dot, and solid curves are the solution of the corrected (uncorrected) uncoupled solver for the first and second RT steps, and their summation, respectively.}
    \label{fig:AnalyticComp}
\end{figure*}

Although the coupled solver provides an accurate solution for the temporal resolution problem, it can be unnecessarily expensive if He and \ce{H2} are included and adding frequency bins accounting for their photo-chemistry. Here we introduce an approximate solution to correct the {\it missing photons} issue with an uncoupled solver, based on the suggestion in \cite{2020MNRAS.499.3594J}.

In Section~\ref{sec:missingPhotonInNeutral}, we have seen that for a neutral cell, the maximum ability to absorb photon is $(1+f_\text{rec})N_\text{\HI}$, where $f_\text{rec}=\Delta t/t_\text{rec}$. \cite{2020MNRAS.499.3594J} proposed a simple method to solve this problem, where they force the number of absorbed photons to be smaller than or equal to the number of atoms that can be ionized. For hydrogen, the number of photons absorbed in a thermochemistry step is
\begin{equation}
\label{equ:jaura}
    \Delta N_{\gamma}=\min{} \left[(1+f_\text{rec})N_\text{\HI}\,, N_{\gamma,0}e^{-\Delta t/t_\text{ion}}\right]\,,
\end{equation}
where $N_\text{\HI}$ and $N_{\gamma,0}$ are the number of neutral hydrogen and the number of hydrogen ionizing photons in a gas cell, respectively.

In the left panel of Figure~\ref{fig:AnalyticComp}, we present the number of photons absorbed in a fully neutral neighbouring cell during the first step after the star ignites, as a function of the size of the time step (with a setup of Test~3, assuming the sizes of all the relative time steps match each other). The solution of the coupled solver (red curve) is regarded as the accurate solution. Compared to this, the uncoupled solver (green curve) deviates from the solution when the time step is larger than $t_\text{ion}$ exhibiting a severe {\it missing photons} issue, while the limited photon absorption correction (equation~\ref{equ:jaura}, blue curve) matches the solution in both large and small $\Delta t$ intervals. 

In Section 5.1.2, we discussed how the difference in cadence between the photon injection and thermochemistry routines can exacerbate the {\it missing photons} issue by repeatedly consuming photons in subsequent RT substeps. As the green lines in Figure~\ref{fig:AnalyticComp} show, if $\Delta t_\star$ is larger than a few $0.01t_\text{rec}$, almost all photons injected will be absorbed in the first RT step, leaving the gas photon-free to recombine in the second RT step. This issue is also solved by the limited photon absorption correction. As shown in the {\it right} panel of Figure~\ref{fig:AnalyticComp}, our correction method ensures that only a small fraction of the total number of photons dumped in an injection step $\Delta t_\star$ are absorbed in the first RT substep. This allows a substantial number of photons to remain in the second substep and suppress gas recombinations. Therefore, the total number of photons absorbed during an injection step shows a perfect agreement with the accurate solution, demonstrating the effectiveness of our correction.

In Figure~\ref{fig:CoupledSolver}, the blue curve shows the momentum feedback obtained with a corrected uncoupled solver, which presents a perfect match to the result of the coupled solver (red curve). Such a perfect correction is based on the fact that the recombination rate, as a function of temperature, is not changing dramatically in our idealized situation. For \HII regions, this holds in most cases because they usually have an equilibrium temperature of about $10^4$\,K, while one should be cautious about other environments with more drastic temperature variations or fluctuations.

\section{Discussion and conclusions}
\label{sec:discussion}
\subsection{When are resolution corrections needed?}
\label{sec:where2correct}
We have demonstrated that resolving the formation phase of \HII regions both spatially and temporally is crucial to couple their momentum and energy feedback to the ISM in a convergent manner. We have also proposed several methods to correct these resolution problems. Here we briefly discuss in what situations we need to include these resolution corrections.

\subsubsection{Spatial resolution issues}
\label{sec:where2spatial}
First of all, we emphasise that eliminating the {\it over-ionization} and {\it over-heating} problems is crucial to obtaining convergent simulation results. Here we will discuss in what situations these two problems become severe numerical issues.

Most RT-coupled Lagrangian codes with similar stellar and gas particle masses $M_{\rm cell}\sim M_{*}\gtrsim10^3\,\Msun$ treat stars born in a GMC as a single stellar particle \citep[e.g.][]{2018MNRAS.480..800H,2019MNRAS.489.4233M}. In this case, the ionizing flux from stellar particles depends only
on the IMF-averaged mass-to-light ratio, which is 
$\langle q \rangle \sim 5\times10^{46}\,\text{photons\,s}^{-1}\Msun^{-1}$ \citep{2015MNRAS.451...34R}. Thus, the resolution effects become important when
\begin{equation}
    n_\text{H} \gtrsim 16\,\text{cm}^{-3}\left(\frac{{\cal R}_i}{10}\right)^{-1}\left(\frac{\langle q\rangle}{5\times10^{46}\,\text{s}^{-1}\Msun^{-1}} \right) \, .
\end{equation}
Regions with higher density in such simulations must be treated with resolution corrections to avoid enhanced ionization feedback.

Higher-resolution simulations with $M_{\rm cell} \lesssim 10\,\Msun$ must treat each massive star as a single star particle \citep[e.g.][]{2019MNRAS.482.1304E}. 
In this case, the density threshold requiring correction becomes
\begin{equation}
    \label{equ:upperdens}
    n_\text{H} \gtrsim 32\,\text{cm}^{-3}\left(\frac{{\cal R}_i}{10}\right)^{-1}\left(\frac{M_\text{cell}}{10\,\Msun}\right)^{-1}Q_{48} \, .
\end{equation}
This result looks similar to that of low-resolution simulations. However, the steep mass-luminosity function of individual massive stars suggests that the \HII regions of low-mass B-type stars with low ionizing photon flux are susceptible to the low-resolution effects. For example, if a $7 \,\Msun$ star is still treated as an individual stellar particle \citep[$Q=8\times10^{43}$\,s$^{-1}$, adopting the fitting function of][]{2002A&A...382...28S}, the density threshold of its Str\"omgren sphere obtained from equation~(\ref{equ:upperdens}) is only $0.0025$\,cm$^{-3}$. Therefore, choosing a reasonable mass floor to conduct explicit RT feedback is a nontrivial problem. Fainter, less-massive stars are also ineffective in providing momentum feedback. Estimated by the early phase analytic results (equation~\ref{equ:H-I} and~\ref{equ:Mi-t}), the total momentum injection of a star follows $p\propto Q^{6/7}t^{3/7}$. For instance, the 7--15\,$\Msun$ stars only provide 1/10 of the momentum feedback compared to the entire 7--120\,$\Msun$ interval (taking the mass-luminosity and mass-lifetime functions of \cite{2002A&A...382...28S} weighted by the \cite{2001MNRAS.322..231K} IMF), while the ionizing rate of a $15 \,\Msun$ can already boost this density threshold to $3$\,cm$^{-3}$. 

Simulations focusing on the formation, accretion, and evolution of individual stars in GMCs have much higher mass resolution and treat stars as sink particles \citep[e.g.][]{2011ApJ...740...74K,2021MNRAS.506.2199G}. The typical mass of gas cells in these simulations is  $M_\text{cell}\sim10^{-3}\,\Msun$, allowing a density threshold $>10^5$\,cm$^{-3}$ following equation~(\ref{equ:upperdens}), so the spatial resolution issues may not be as problematic for these simulations. However, these simulations usually aim to resolve the physics in ultra-dense regions of $n>10^7$\,cm$^{-3}$, where we still need to carefully deal with similar issues for the low-mass ionizing stars.

In summary, the initial Str\"omgren sphere can only be spatially resolved in regions with relatively low density or around a particularly strong ionizing source. Therefore, in simulations with large dynamic ranges, such as galaxy formation, multiphase ISM, and GMC, it becomes essential to consider implementing a correction to prevent divergent feedback from unresolved \HII regions.

\subsubsection{Temporal resolution issues}
The requirement to ignore the temporal resolution correction is much more stringent. 
As presented in Appendix~\ref{sec:step-by-step}, if $\Delta t_\text{RT}\gg t_\text{rec}$ the {\it missing photons} issue will appear as long as the spatial resolution is high (i.e. ${\cal R}_i\gg100$, equation~\ref{equ:Kcrit}) so that the number of H atoms in a cell is much less than the number of photons. Failure to meet this requirement is not problematic because the {\it missing photons} issue only happens when the radiation meets neutral gas unless there are differences between the sizes of the photon injection step and the thermochemistry step.

When the injection cadence is not the same as thermochemistry, the simplified time-stepping criterion of photon injection step $\Delta t_\star\lesssim 0.1 t_\text{rec}N_\text{sub}/(N_\text{sub}-1)$ implies a density threshold of (combining equations~\ref{equ:trec} and \ref{equ:tC})
\begin{equation}
    n_\text{H}\gtrsim 0.004\,\text{cm}^{-3}\,\eta^{-3}\left(\frac{N_\text{sub}}{N_\text{sub}-1}\right)^3\left(\frac{\tilde{c}}{10^{-3}c}\right)^3\left(\frac{{\cal R}_i}{10}\right)Q_{48}^{-1}\, .
\end{equation}
This value is too small for most classes of simulations unless the spatial resolution is extremely high. Thus, we strongly suggest that all RT implementations with different sourcing and absorption cadences should consider adopting a special time-stepping scheme or a temporal resolution correction to avoid the {\it missing photons} issue. If the correction is not implemented, choosing a smaller Courant factor $\eta$ or faster speed of light $\tilde{c}$ is the easiest way to significantly increase the density threshold ($\sim \eta^{-3} \tilde{c}^3$) and ensure the correct simulation behaviour. 

\subsection{\ce{H2} and helium}

In this work, we mainly focused on pure hydrogen gas, while the thermochemistry of molecular hydrogen and helium can also play a significant role in realistic astrophysical environments.

\ce{H2} in the \HII regions can be converted to H atoms through photodissociation (PD) by 11.2--13.6\,eV Lyman-Werner (LW) band photons and photoionization by $>15.2$\,eV UV photons. Theoretically, for typical massive stars, the PD-front is merged with the I-front in the early phase, i.e. the \ce{H2} molecules are dissociated to H atoms and ionized to \ce{H+} ions almost simultaneously \citep{1996ApJ...458..222B}. On the other hand, the processes forming \ce{H2} (dust catalysis and gas phase formation) have very low coefficients compared to the dissociation rates \citep{2018MNRAS.479.3206N}. For example, comparing the ratio between the rate of the fastest \ce{H2} formation channel, the dust catalyst rate $\alpha^\text{D}_\text{\ce{H2}}$, and LW band dissociation rate $\sigma^\text{LW}_\text{\ce{H2}}$ with that of the ionization and recombination rates of hydrogen, we have $(\alpha^\text{D}_\text{\ce{H2}}/\sigma^\text{LW}_\text{\ce{H2}})/(\alpha_\text{\HII}\sigma_\text{\HI}) \approx0.0001\,$ \citep[assume a Milky-way dust-to-gas ratio of $\sim0.01$,][]{2007ApJ...663..866D}. Thus, both spatially and temporally resolving the dissociation of \ce{H2} when the \HII region is forming is much easier than doing this for H atoms.

For helium, the case B radiative recombination rate for \HeII is similar to \HI \citep[$2.72\times10^{-13}T_4^{-0.789}$\,cm$^3$\,s${^{-1}}$,][]{1999ApJ...514..307B}. Adopting a primordial abundance of $X=0.76$, the number density ratio of He to H is 0.08, so the initial Str\"omgren mass of helium is about $3.5Q_\text{He}/Q_\text{H}$ times as much as that of hydrogen, where $Q_\text{H}$ and $Q_\text{He}$ are the ionization photon rates of H and He, respectively (equation~\ref{equ:HeIIMass}). Therefore, for a given star, the mass resolution of its helium Str\"omgren sphere is $15Q_\text{He}/Q_\text{H}$ times that of hydrogen. Resolving the initial Str\"omgren mass of helium is even easier for most O-type stars with $Q_{\rm He}>Q_\text{H}/15$ so long as their $T_\text{eff}\gtrsim3.5\times10^4$\,K \citep[][]{2003ApJ...599.1333S,2005A&A...436.1049M}, so the mass resolution for helium can meet this requirement in most situations as long as hydrogen ionization is spatially resolved.

The ionization of \HeII requires photons with $h\nu>54.4$\,eV. $Q_\text{\HeII}$ is usually very small for ordinary massive stars and the recombination rate for \HeIII is about 10 times larger \citep[$2.19\times10^{-12}\,\text{cm}^{-3}$\,s$^{-1}$][]{1996ApJS..103..467V}. This makes the \HeIII Str\"omgren sphere with very little mass difficult to be resolved. Additionally, resolving or correcting the \HeII ionization needs to obtain the correct \HeII fraction at first, which is not always guaranteed. Thus, the mass and physical states of \HeIII content might be highly uncertain in most simulations but overall \HeIII has a negligible contribution to the feedback.

For the temporal resolution of helium, the characteristic time-scale should be similar to that of hydrogen, the recombination time-scale $\{t_\text{rec,\HeII}, t_\text{rec,\HeIII}\}=\{1/\alpha_\text{\HeII}(n_\text{H}+n_\text{He}),1/\alpha_\text{\HeIII}(n_\text{H}+2 n_\text{He})\}$. As the recombination coefficient ($\alpha_\text{\HeII}$) of \HeII is comparable to that of H, $t_\text{rec,\HeII}$ is also comparable to that of H, making it similarly difficult to be temporally resolved. However, for \HeIII, $\alpha_\text{\HeIII}$ is one order of magnitude larger, making $t_\text{rec,\HeIII}$ one order of magnitude smaller. This, in turn, makes the \HeII--\HeIII thermochemistry more challenging to be temporally resolved. 

In summary, resolving the hydrogen Str\"omgren sphere presents the greatest numerical challenge and dynamical significance compared to \ce{H2} and \HeI. As such, the resolution requirements and correction methods proposed in this paper remain relevant even in more complex chemical networks. For primordial chemical networks, we outline the implementation of our spatial resolution correction scheme, which includes He and \ce{H2} chemistry, in Appendix~\ref{sec:fcor}. Additionally, readers can refer to \cite{2020MNRAS.499.3594J} for a limited absorption correction that also considers He and \ce{H2}.

\subsection{\HII regions in inhomogeneous density structures}
Realistic MCs exhibit highly filamentary structures, where protostars form in overdensities characterized by power-law density profiles ($\rho\propto r^{-w}$, \citealt{1981MNRAS.194..809L,2010ApJ...716..433K}, see \citealt{2014prpl.conf...27A} for a review). 
\cite{1990ApJ...349..126F} demonstrated that \HII regions in media with $w>3/2$ can expand in an accelerating manner, driving a ``champagne flow". In complex structures, \HII regions tend to expand toward the rarefied regions and stall in the dense regions as blister-type regions \citep{2006ApJ...647..397M,2012ApJ...745..158G,2019MNRAS.487.2200Z,2022ApJ...934L...8J}. 

In numerical simulations, these effects can be self-consistently modelled on a scale larger than the numerical resolution, but the small-scale density gradients are smoothed due to the inadequate resolution. In the absence of a sub-grid turbulence model, gas represented by a cell is assumed to be uniform. Consequently, numerically unresolved \HII regions suffer from {\it over-ionization} and {\it over-heating} issues similar to those in the uniform medium, while our corrections described in Section~\ref{sec:Rifix} ensure the correct ionized mass predicted by the uniform solutions. However, the Strömgren mass evaluated based on the mean density may overestimate the ionized mass since stars are embedded in unresolved overdense cores where the \HII regions are trapped by the high density. This potentially leads to an overestimation of feedback from these trapped \HII regions.

Nonetheless, once the \HII regions break through
these dense cores due to expansion or the assistance of stellar winds and radiation pressure \citep[e.g.][]{2020MNRAS.492..915G,2021MNRAS.501.1352G}, they are easier to resolve (${\cal R}_i\propto n^{-1}$) and the local density fluctuations will have a less significant impact on the global properties of the \HII regions as they expand toward the low-density gas.
High-resolution radiation-hydrodynamic simulations by \cite{2006ApJ...647..397M} showed that though the expansion of \HII region in turbulent MCs is slower than that in uniform medium with the same mean density due to the overdensity around the star,  over time, the mean radius can eventually catch up with the uniform case. Hence, our corrections are considered reasonable approximations to capture the feedback from unresolved \HII regions in realistic environments.

\subsection{Other early feedback channels and their combination}
In addition to ionization feedback, luminous massive stars also provide energetic early (pre-SN) feedback through radiation pressure and high-velocity stellar winds.
However, resolving the effects of radiation pressure in numerical simulations presents its own challenges due to the high spatial and temporal resolution requirements, especially in dense dusty regions \citep[e.g.][]{2018MNRAS.480.3468K,2019MNRAS.483.4187H}. Similarly, modelling the effects of stellar winds also requires high resolution simulations \citep[e.g.][]{2021MNRAS.508.1768P}. The rapid energy loss due to the turbulent mixing at the wind bubble/shell interface poses additional challenges that need to be addressed to accurately simulate the wind feedback \citep{2021ApJ...914...89L,2021ApJ...914...90L}. The combination of all these early feedback channels in simulations is nonlinear and can provide a complicated collective impact rather than their simple superposition \citep[e.g.][]{2018MNRAS.478.4799H,2021MNRAS.501.1352G}. Additional challenges on the numerical resolution can be introduced by such physics 
and requires further investigation to better understand their collective impact on galaxy formation simulations.

\subsection{Summary}
\label{sec:conclusion}
We have performed a suite of radiation hydrodynamic simulations of idealized \HII regions with the moment-based M1 closure RT solver in the moving-mesh code \arepo. Below, we list our main findings.
\begin{enumerate}
    \item Sufficient mass (spatial) resolution (${\cal R}_i=M_\text{S}/M_\text{cell}$) is critical to accurately capture the evolution and feedback of simulated \HII regions. With a spatial resolution higher than $\mathcal{R}_i = 10$ (more than 10 cells inside the initial Str\"omgren sphere), the momentum and energy feedback from an expanding \HII region can be reproduced with an acceptable numerical error. If we want to reproduce the profiles of shock parameters, the resolution should be higher than $\mathcal{R}_i = 100$.
    
    \item In the formation phase of \HII regions, insufficient spatial resolution lowers the ionization fraction of gas cells but can lead to an overestimation of the ionized mass when the \HII region is forming, i.e. the {\it over-ionization} problem. In the expansion phase, the insufficient spatial resolution will fail to heat the ionized gas cells to a correct temperature and result in insufficient cooling, the total thermal energy of gas will thus also be overestimated and lead to enhanced momentum feedback, i.e. the {\it over-heating} problem. It is crucial to avoid {\it over-ionization} and {\it over-heating} problems in simulations because they both lead to divergent feedback when the \HII regions are unresolved. To correct the spatial resolution problems, we can consider both lowering the number of neighbours $N_\text{nb}$, as well as introducing a correction to reestablish the accurate balance between ionization and recombination. However, although these corrections can correct the momentum feedback from \HII regions, they may not help enough for dynamic multi-phase gas structures. The {\it over-heating} problem will turn the highly-ionized, hot ($>10^4$\,K) gas concentrated in the \HII regions into diffuse, partially ionized, warm ($\sim8000$\,K) gas, changing the multi-phase gas structure of the ISM in star-forming regions. 
    
    \item If the size of the time step is too large, almost all
    photons dumped in the first step will be lost and there is no mechanism to compensate for these missing photons, leading to the {\it missing photons} issue. Most simulations of real astrophysical systems (e.g., galaxy formation, multi-phase ISM, GMC) cannot temporally resolve the ionization timescale of the gas. Thus, the {\it missing photons} issue should be treated carefully if the cadences of photon injection and thermochemistry are different in these simulations. A rough time-stepping criterion for the injection time step is $\Delta t_\star\lesssim0.1t_\text{rec}N_\text{sub}/(N_\text{sub}-1)$ when each injection time step is associated with $N_\text{sub}$ RT steps. 
    
    \item If the ratio of injection step size to RT step size ($t_\star/t_\text{RT}$) is too large, the {\it missing photons} issue will repeat after each injection. Ionizing fluxes reduced in this way will be insufficient to ionize and heat the gas and reduce the feedback from the \HII regions. Using a thermochemistry solver coupled with the photon densities can obtain an accurate solution and solve the {\it missing photons} problem fundamentally, but this is an expensive choice when including many species and frequency bins. Alternatively, the uncoupled solver with the limited photon absorption approximation proposed by \cite{2020MNRAS.499.3594J} is a much cheaper but very powerful correction for this issue.
    
    \item Spatially and temporally resolving the thermochemistry of hydrogen is generally the most challenging part compared to that of \ce{H2} and \HeI. While resolving and correcting the \HeII--\HeIII thermochemistry also presents its own difficulties, it is of relatively lesser importance for stellar feedback. Once the ionization feedback from hydrogen is spatial and temporally resolved, those for \ce{H2}, \HeI will be already solved, alleviating the complexity when designing resolved ionization feedback models. We also outline a method to implement our spatial resolution correction scheme including He and \ce{H2} chemistry in Appendix~\ref{sec:fcor}, and we refer the readers to \cite{2020MNRAS.499.3594J} for the limited absorption correction including He and \ce{H2}.
\end{enumerate}

\section*{Acknowledgements}

We thank Volker Springel for giving us access to \arepo. YD is grateful to Josh Borrow and Yang Ni for useful discussions. YD was a visiting student at the Massachusetts Institute of Technology sponsored by the ZhengGang Fund of NJU through the ZhengGang Scholarship for Overseas Study. AS acknowledges support under an Institute for Theory and Computation Fellowships at the Center for Astrophysics $\vert$ Harvard \& Smithsonian. MV acknowledges support through NASA ATP 19-ATP19-0019, 19-ATP19-0020, 19-ATP19-0167, and NSF grants AST-1814053, AST-1814259, AST-1909831, AST-2007355 and AST-2107724. GLB acknowledges support from the NSF (AST-2108470, XSEDE grant MCA06N030), NASA TCAN award 80NSSC21K1053, and the Simons Foundation through their support of the Learning the Universe collaboration. The simulations of this work were run on the MIT Engaging cluster, the Anvil cluster at Purdue University as part of XSEDE through TG-PHY220025, and the Stampede2 HPC resource at Texas Advanced Computing Center as part of XSEDE through TG-AST200007 and TG-MCA06N030. We use \textsc{python} packages {\sc NumPy} \citep{harris2020array}, {\sc SciPy} \citep{2020SciPy-NMeth}, {\sc astropy} \citep{2013A&A...558A..33A,2018AJ....156..123A}, and {\sc matplotlib} \citep{Hunter:2007} to analyze and visualize the simulation data.

\section*{data availability}
The data that support the findings of this study are available from the corresponding author, upon reasonable request.



\bibliographystyle{mnras}
\bibliography{IonizationFeedback} % if your bibtex file is called example.bib

\appendix
\section{Cooling Rates}
\label{sec:cooling}
In this section, we list the fitting formulas accounting for the cooling processes in pure hydrogen \HII regions used in this work \citep{1992ApJS...78..341C}. 

The total cooling rate can be split as follows
\begin{equation}
    \Lambda=\Lambda_\text{rec}+\Lambda_\text{bb}+\Lambda_\text{bf}+\Lambda_\text{ff}\,,
\end{equation}
where $\Lambda_\text{rec}$, $\Lambda_\text{bb}$, $\Lambda_\text{bf}$, and $\Lambda_\text{ff}$ are the cooling rates accounting for recombination, bound-bound collision, bound-free, collision, and free-free collision (bremsstrahlung), respectively
We adopt the following approximation results 
 \begin{equation}
\Lambda_\text{rec}=8.7\times10^{-27}n_en(\text{\HII})T^{1/2}\frac{T_3^{-0.2}}{1+T_6^{0.7}},
\label{equ:H_rec}
 \end{equation}

 \begin{equation}
\Lambda_{\mathrm{ff}}=1.42\times10^{-27}T^{1/2}g_{\mathrm{ff}}n_en_\text{\HII}.
\label{equ:ff}
 \end{equation}
 
 \begin{equation}
\Lambda_\text{bb}=7.5\times10^{-19}n_en(\text{\HI})\frac{e^{-\frac{\epsilon_{21,1}}{k_\text{B}T}}}{1+T_5^{1/2}},
\label{equ:H_bb}
\end{equation}
\begin{equation}
\Lambda_\text{bf}=1.27\times10^{-21}n_en(\text{\HI})T^{1/2}\frac{e^{-\frac{I_1}{k_\text{B}T}}}{1+T_5^{1/2}}.
\label{equ:H_bf}
 \end{equation}
where $k_B$ is the Boltzmann constant, $I_1=13.6$\,eV is the ionization potential of H, $\epsilon_{21}=10.2$\,eV is the energy level difference between the ground state and first excited state of the hydrogen atom.

\section{Step-by-step tracking of the missing photons issue}
\label{sec:step-by-step}
\begin{figure*}
	\includegraphics[width=2\columnwidth]{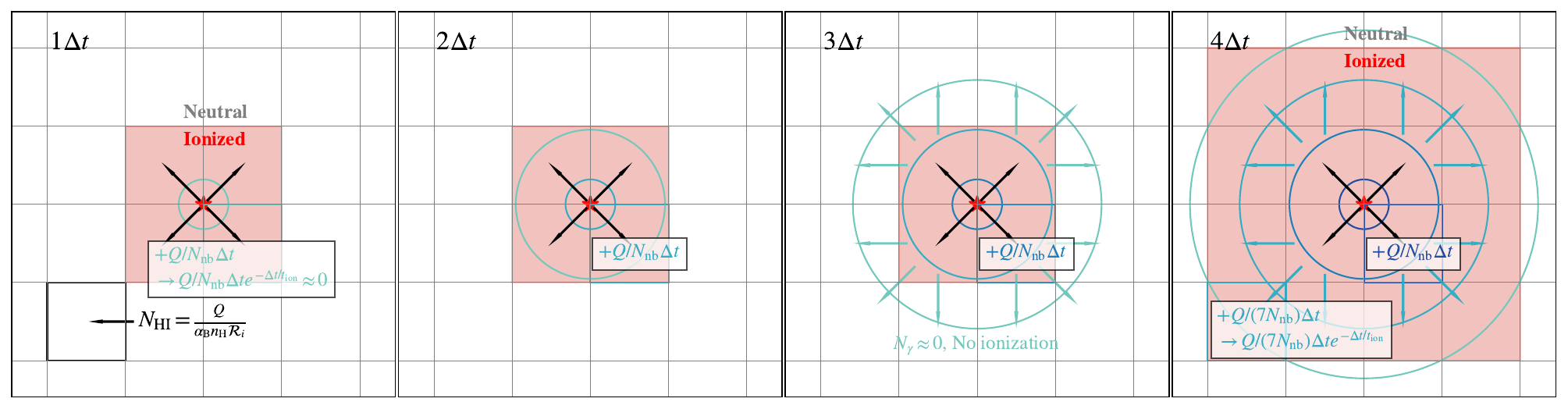}
    \caption{Cartoon illustrating the step-by-step photon propagation and thermochemistry after a massive star ignites. {\it Far left}: The 1st step ($t=0\rightarrow 1\Delta t$), when the star ignites and dumps $Q\Delta t$ photons into its $N_\text{nb}$ neighbouring cells. The neighbouring cells are all fully ionized and the {\it missing photons} issue happens for the first time. {\it Middle left}: The 2nd step ($t=1\Delta t\rightarrow 2\Delta t$), when the photons stay in these neighbouring cells and the star dumps $Q\Delta t$ photons again. The photon density in these cells begins to increase linearly. {\it Middle right}: The 3rd step ($t=2\Delta t\rightarrow 3\Delta t$), when the photons dumped in the first step will propagate outwards into the cells enclosing the $N_\text{nb}$ neighbouring cells but do not lead to ionization because most of them are absorbed as {\it missing photons}. {\it Far right}: The 4th step ($t=3\Delta t\rightarrow 4\Delta t$), when the photons dumped in the 2nd step now propagate into the cells enclosing the nearest neighbouring cells and the second {\it missing photons} issue happens.}
    \label{fig:missingphoton}
\end{figure*}
In this section, we demonstrate the {\it missing photons} issue by step-by-step tracking the evolution of photon density immediately after the star ignition with the aid of Figure~\ref{fig:missingphoton}.  

Like previous tests, a stellar particle with ionizing photon rate $Q$ sits at the centre of a pure hydrogen uniform simulation box which we take to be a regular Cartesian mesh here for convenience. Once the star ignites, $Q\Delta t$ photons are dumped into its $N_\text{nb}$ neighbouring cells isotropically at the beginning of each step and then propagate outward. After photon dumping and propagating, the code will perform the thermochemistry and photon absorption by solving equation~(\ref{equ:thermoeq}) and equation~(\ref{equ:photonchange}). 
We simply assume the photon dumping, propagating, and thermochemistry is conducted successively during one integration step, meaning that the performance of these steps shares the same cadence.

{\bf The 1st step ($t=0\rightarrow 1\Delta t$):} Once the star ignites, $N_{\gamma}=Q\Delta t$ photons are dumped into the neighbouring cells during the first $\Delta t$. As described in Section~\ref{sec:dumping}, these photons are allocated to $N_{\rm nb}$ cells in the neighbourhood list found by the neighbour search. The other cells are physically impossible to receive any photons because $\Delta t$ must be smaller than the Courant criterion $\Delta t_{\rm C}$ so that no photons can escape from these neighbouring cells (assuming the neighbouring cells can completely wrap the centre star). Thus, the number of photons in each neighbouring cell (at $t=\Delta t$) is 
\begin{equation}
    N_{\gamma,1} = \frac{Q}{N_{\rm nb}} \Delta t \, ,
\end{equation}
where the subscript of $N_\gamma$ accounts for the step.

The number of photons consumed in this step is (equation~\ref{equ:photonchangeS1})
\begin{equation}
    \Delta N_{\gamma,1} = \frac{Q}{N_{\rm nb}}\Delta t(1-e^{-\Delta t/t_{\rm ion}})\approx \frac{Q}{N_{\rm nb}}\Delta t \, ,
\end{equation}

which exhibits that almost all photons dumped into the gas will be consumed when the $t_\text{ion}$ (or MFP) is unresolved.
Again, we emphasize that the neutral density $n_{\rm HI}$ involved here is taken as the value at the beginning of the step (i.e. $1\cdot n_{\rm H}$) because the photon absorption equation is not coupled with the thermochemistry equations.

On the other hand, the total number of neutral hydrogen atoms in a cell is
\begin{equation}
    \label{equ:NHI}
    N_\text{\HI} = \frac{M_{\rm cell}}{m_{\rm H}}=\frac{Q}{\alpha_{\rm B}n_{\rm H}{\cal R}_i} \, .
\end{equation}
Assuming one photon can ionize one neutral hydrogen atom (MFP unresolved), in this case, if $\Delta N_{\gamma}>(1+f_\text{rec})N_\text{\HI}$, then $\Delta N_{\gamma}$ photons will be absorbed but only $N_\text{\HI}$ atoms can be ionized. Here $f_\text{rec}$ is a factor accounting for the recombination of ions. In other words, $\delta N_{\gamma}$ photons simply disappear without any effect on the gas state. The photon number is no longer conserved. For each neighbouring cell, the number of missing photons is
\begin{equation}
    \label{equ:dN1}
    \delta N_{\gamma,1} = \frac{Q}{N_{\rm nb}}\Delta t-(1+f_\text{rec})\frac{Q}{\alpha_{\rm B}n_{\rm H}{\cal R}_i} \, ,
\end{equation}
where $f_\text{rec}$ can be estimated as $\Delta t/t_\text{rec}$ because $\Delta t\sim t_\text{rec}\gg t_\text{ion}$.

On the contrary, if the MFP is still unresolved but $\Delta N_{\gamma}<N_\text{\HI}$, the ionization will be overestimated in this step. Nonetheless, this error can be relieved in the subsequent steps by the shifting of ionization and recombination rates. Thus, it has minor effects on the results compared to the {\it missing photons} issue.

{\bf The 2nd step ($t=1\Delta t\rightarrow 2\Delta t$):} If the {\it missing photons} issue happened in the first step, the neighbouring cells must be fully ionized in this step ($\Delta N_{\gamma}>N_\text{\HI}$). With $\Delta t< \Delta t_{\rm C}$, the photons will stay in these neighbouring cells during the next few steps as long as $t<\Delta x/\tilde{c}=\Delta t_{\rm C}/\eta$, where $\eta$ is the Courant factor. Again, the change of photon density due to thermochemistry is calculated by equation~(\ref{equ:photonchange}) with a zero $n_{\rm HI}$ because the cell is fully ionized. Thus, the photon number density will increase linearly without any {\it missing photons} issues in these steps.

Here we define $M=\lceil \Delta t_{\rm C}/\eta \Delta t \rceil$, the period of this linear increasing stage is thus $M\Delta t$. After the $M$th step, the photons dumped in the first step will propagate outwards into the cells enclosing the $N_{\rm nb}$ neighbouring cells. For convenience, we simply assume $M=3$ (i.e. $2<\Delta t_{\rm C}/\eta \Delta t<3$).

{\bf The 3rd step ($t=2\Delta t\rightarrow 3\Delta t$):} In the third step, the photons dumped in the first step will propagate outwards into the cells enclosing the $N_{\rm nb}$ neighbouring cells. However, the photons dumped in the first step have all been absorbed and become missing photons. Thus, what actually has an impact on the enclosing cells is the photons dumped in the 2nd step, which will arrive in these cells in the next step. 

{\bf The 4th step ($t=3\Delta t\rightarrow 4\Delta t$):} The photons dumped in the 2nd step now propagate into the cells enclosing the $N_{\rm nb}$ nearest neighbouring cells. These photons are dumped in the linear increasing stage so they suffer no reaction and keep the number of $Q\Delta t$ before they begin to ionize these cells. Once the ionization is conducted, the {\it missing photons} issue will happen in a similar manner as in the 1st step, but now $Q\Delta t$ photons are propagating into $N_{\rm shell}>N_{\rm nb}$ cells. We can simply estimate $N_{\rm shell}$ as 
\begin{equation}
   N_{\rm shell} =\frac{\left[V(2c\Delta t) - V(c\Delta t)\right]}{V(c\Delta t)} = 7{N_{\rm nb}} \, .
\end{equation}
Thus, the number of photons consumed by a cell in this step is
\begin{equation}
    \Delta N_{\gamma,4} = \frac{Q}{7N_{\rm nb}}\Delta t(1-e^{-\Delta t/t_{\rm ion}})\approx  \frac{Q}{7N_{\rm nb}}\Delta t\, .
\end{equation}
The total number of neutral hydrogen atoms in a cell is the same as Eq.~(\ref{equ:NHI}) and the missing photons issue is alleviated when it happens again
\begin{equation}
    \delta N_{\gamma,4}= \frac{Q}{7N_{\rm nb}}\Delta t-(1+f_\text{rec})\frac{Q}{\alpha_{\rm B}n_{\rm H}{\cal R}_i} \, .
\end{equation}
In general, after the occurrence of the second {\it missing photons} issue, the photon density begins to increase linearly again before the $(2M+2)$th step. In the $(2M+2)$th step the third photon-missing event will happen and in the $K(M+1)$th step the {\it missing photons} issue will happen $K$ times. The number of cells involved in the $K$th {\it missing photons} issue is 
\begin{equation}
    N_{\rm shell} = \frac{4\pi (K\tilde{c}\Delta t)^3/3-4\pi \left[(K-1)\tilde{c}\Delta t\right]^3/3}{4\pi (\tilde{c}\Delta t)^3/3}{N_{\rm nb}} \approx3K^2{N_{\rm nb}} \, .
\end{equation}
The {\it missing photons} issue will keep happening until $\delta N^{K(M+1)}_{\gamma}\leq0$, i.e.
\begin{equation}
\label{equ:missingK}
        K \approx\sqrt{\frac{{\cal R}_i}{3N_{\rm nb}(1+\Delta t/t_\text{rec})}\left(\frac{\Delta t}{t_{\rm rec}}\right)} \, .
\end{equation}
To strictly avoid the occurrence of {\it missing photons} issue, $K$ should be $\leq1$ and we have equation~(\ref{equ:Kcrit}) (drop the factor 3). 

\section{Validation of time-step criterion}
\label{sec:validate}
In Section~\ref{sec:Subcycling}, we proposed a time-stepping criterion, $\Delta t_\star\lesssim0.1t_\text{rec}N_\text{sub}/(N_\text{sub}-1)$, which avoids oscillations in the \HII fraction and repeated {\it missing photons} when photon injection and thermochemistry are conducted with different cadences. The strict derivation of this criterion needs solving equation~(\ref{equ:dN2}) to find out when $\delta N_{\gamma,2}<0$. This is extremely complicated and unnecessary. Alternatively, here we validate this time-stepping criterion by discussing the solution of this equation with several sets of parameters.

To avoid the oscillation of \HII fraction, we do not need strict $\delta N_{\gamma,2}<0$ but only a significant fraction of photons survived when the {\it missing photons} issue happens again because of recombination (equation~\ref{equ:dN2}) to suppress the recombination in the next substep. 

\begin{figure}
	\includegraphics[width=\columnwidth]{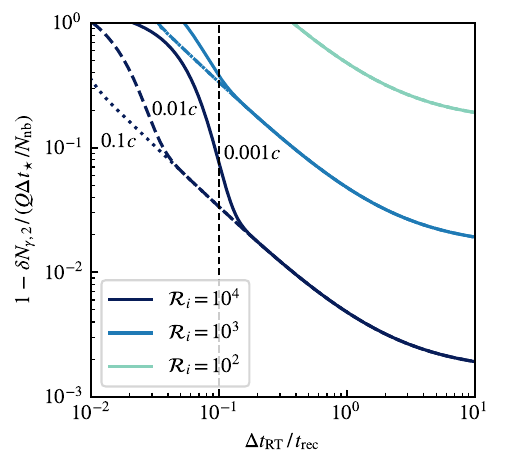}
    \caption{Fraction of survived photons (normalized to $Q\Delta t_\star/N_\text{nb}$) as a function of $\Delta t_\text{RT}$ assuming $N_\text{sub}=2$. Curves with different colours are obtained with different spatial resolutions ${\cal R}_i={10^2,10^3,10^4}$. Curves marked solid, dash, and dot lines are obtained with different speeds of light ${\tilde{c}}_i=\{0.1,0.01,0.001\}c$.} 
    \label{fig:criterion}
\end{figure}

In Figure~\ref{fig:criterion}, we present the fraction of survived photons (normalized to $Q\Delta t_\star/N_\text{nb}$) as a function of $\Delta t_\text{sub}$ under different spatial resolutions and speeds of light assuming $N_\text{sub}=2$. With $\Delta t_\text{RT}<0.1$ ($\Delta t_\star<0.2$) most runs can leave $>10$ per cent photons after the {\it missing photons} issue, except the ${\cal R}_i=10^4$ runs with faster speeds of light. The low spatial resolution cases show weak dependence on the speed of light because there are always enough H atoms compared with the photon injection rate. In galaxy formation and multiphase ISM simulations, single \HII regions can hardly reach such a high spatial resolution as ${\cal R}_i=10^4$, we thus argue that our $\Delta t_\star\lesssim0.1t_\text{rec}N_\text{sub}/(N_\text{sub}-1)$ criterion is reasonable and robust for most cases.

\section{Including \ce{H2} and He chemistry}
\label{sec:fcor}
To implement our spatial resolution correction scheme including \ce{H2} and He chemistry, we need to modify the parameters in the correction factor (equation \ref{equ:f_cor}) for different species. Following Section~\ref{sec:resolutionDef}, the mass resolution ${\cal{R}}_i$ refers to how many Lagrangian resolution elements there are to resolve the initial Str\"omgren mass, we can modify equation~(\ref{equ:resolution}) to
\begin{equation}
    \mathcal{R}^\text{\HII}_{i,k} = \frac{M_\text{S,\HII}}{XM_\text{cell,k}}\, ,
\end{equation} 
where $M_\text{S,\HII}$ is the theoretical mass of hydrogen Str\"omgren sphere, $X$ is the hydrogen mass fraction.

For an \HII Str\"omgren sphere, we still assume $x_\text{\HII}=x_e={\cal R}^\text{\HII}_{i,k}$ and we have
\begin{equation}
    f^\text{\HI}_\text{cor}= \frac{\alpha_\text{\HII} n_\text{H}}{\tilde{c}\sum_\text{\HI}\sigma_{i\text{\HI}}n^i_{\gamma}}\left({\cal R}^\text{\HII}_{i,k}\right)^2 \, .
\end{equation}

For \HeII Str\"omgren sphere, we temporarily ignore the conversion between \HeII and \HeIII,  $x_{\text{\HeII}+\text{\HeIII}}=y{\cal R}^\text{\HeII}_{i,k}$, $x_e = {\cal R}^\text{\HII}_{i,k}+y{\cal R}^\text{\HeII}_{i,k}$, where ${\cal R}^\text{\HII}_{i,k}=M_\text{S,\HeII}/(1-X)M_{\text{cell},k}$, $y=(1-X)/4X$
\begin{equation}
\label{equ:HeIIMass}
    M_\text{S,\HeII}+M_\text{S,\HeIII} \approx \frac{m_\text{He}Q_\text{\HeII}}{ (n_\text{H}+n_\text{He})\alpha_\text{\HeII}}\,.
\end{equation}
The correction factor for \HeII is thus,
\begin{equation}
    f^\text{\HeII}_\text{cor}= \frac{\alpha_\text{\HeII} n_\text{H}}{y\tilde{c}\sum_\text{\HeI}\sigma_{i\text{\HeI}}n^i_{\gamma}}\left({{\cal R}^\text{\HII}_{i,k}}+{\cal R}^\text{\HeII}_{i,k}\right){\cal R}^\text{\HeII}_{i,k}\, .
\end{equation}

For \HeIII, each doubly ionized helium contributes two electrons, similarly, we have
\begin{equation}
    M_\text{S,\HeIII} = \frac{m_\text{He}Q_\text{\HeIII}}{ (n_\text{H}+2n_\text{He})\alpha_\text{\HeIII}}\,,
\end{equation}
and the correction factor
\begin{equation}
    f^\text{\HeIII}_\text{cor}= \frac{\alpha_\text{\HeIII} n_\text{H}}{y\tilde{c}\sum_\text{\HeII}\sigma_{i\text{\HeIII}}n^i_{\gamma}}\left({\cal R}^\text{\HII}_{i,k}+2{\cal R}^\text{\HeIII}_{i,k}\right){\cal R}^\text{\HeIII}_{i,k} \, .
\end{equation}

Notice that the correction of \HeIII should be based on resolved \HeII content or the \HeIII usually suffers an under-ionization. However, our
correction can still ensure the numerical convergence by setting an upper limit of \HeIII mass.

Both photoionization and photodissociation can destroy \ce{H2} molecules. However, the $>15.2$\,eV \ce{H2} ionizing photons can also ionize atomic H so their opacity in ISM is similar to the atomic Str\"omgren radius. On the other hand, LW band photons can propagate far away and dissociate more \ce{H2}, leading to a ``molecular St\"omgren sphere'' filled with an \HI envelope and an \HII region. Thus, we treat the molecular St\"omgren sphere in a similar we to the \HeII Str\"omgren sphere, i.e., ignore the conversion between \HI and \HII inside $n_{\text{\HI}+\text{\HII}}={\cal R}^\text{\ce{H2}}_{i,k}$ and 
\begin{equation} M_\text{S,\HI}+M_\text{S,\HII}=\frac{m_\text{H}Q_\text{LW}}{\alpha_\text{\ce{H2}}n_\text{H}}+M_\text{S,\HII}
\end{equation}
where the formation rate $\alpha_\text{\ce{H2}}$ is usually dominated by the  dust-catalyzed formation process $\alpha^\text{D}_\text{\ce{H2}}$. We have

\begin{equation}
    f^\text{\ce{H2}}_\text{cor}= \frac{\alpha^\text{D}_\text{\ce{H2}} n_\text{H}}{\tilde{c}\sigma^\text{LW}_\text{\ce{H2}}n^\text{LW}_{\gamma}}{\cal R}^\text{\ce{H2}}_{i,k} \, .
\end{equation}

\begin{figure}
	\includegraphics[width=\columnwidth]{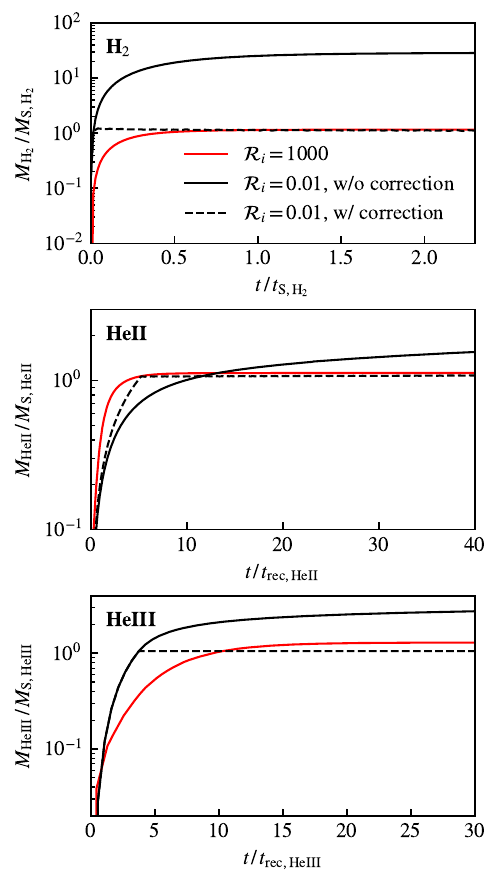}
    \caption{Evolution of \HI ({\it top}), \HeII ({\it middle}), and \HeIII ({\it bottom}) masses in a $\mathcal{R}_i=0.01$ resolution with (black dash curves) and without (black solid curves) the spatial resolution correction, compared to the $\mathcal{R}_i=1000$ (red solid curves) solution. Our correction helps to correct the masses of all species, as seen by their convergence towards the expected analytic results and those run with the $\mathcal{R}_i=1000$ resolution.}
    \label{fig:multifreq}
\end{figure}

We conducted a suite of runs to test the effectiveness of our corrections for \ce{H2} and helium, which were set up similar to Test~2 (Section~\ref{sec:EXPtest}). We simulate the R-type expansion of the Str\"omgren spheres for different species to demonstrate that the $\mathcal{R}_i$ fix can correct the species masses to the theoretical value (Str\"omgren mass) in the early phase. In Figure~\ref{fig:multifreq}, we present the evolution of \HI ({\it top}), \HeII ({\it middle}), and \HeIII ({\it bottom}) masses in a $\mathcal{R}_i=0.01$ resolution with (black dash curves) and without (black solid curves) the $\mathcal{R}_i$ fix. We also plot the $\mathcal{R}_i=1000$ (red solid curves) solution as a comparison. Similar to the hydrogen case (Figure~\ref{fig:RiFixT1}), our results show that the masses of all species can be constrained close to their expected analytic result.

%\begin{thebibliography}{99}
%\end{thebibliography}

%%%%%%%%%%%%%%%%%%%%%%%%%%%%%%%%%%%%%%%%%%%%%%%%%%

% Don't change these lines
\bsp	% typesetting comment
\label{lastpage}
\end{document}